\documentclass[12pt, draftclsnofoot, onecolumn]{IEEEtran}
\usepackage{amsmath,amsfonts}
\usepackage{amssymb}
\usepackage{mathtools}
\usepackage{mathrsfs}
\usepackage{bm}
\usepackage{braket}
\usepackage{algorithmic}
\usepackage{algorithm}
\usepackage{array}
\usepackage{microtype}
\usepackage{textcomp}
\usepackage{stfloats}
\usepackage{url}
\usepackage{verbatim}
\usepackage{graphicx}
\usepackage{cite}
\usepackage{xcolor}
\usepackage{xfrac}
\usepackage[utf8]{inputenc}
\usepackage[T1]{fontenc}
\usepackage{balance}
\usepackage{multirow}
\usepackage{booktabs}
\usepackage{optidef}
\usepackage[caption=false,font=normalsize,labelfont=rm,textfont=rm]{subfig}
\def\BibTeX{{\rm B\kern-.05em{\sc i\kern-.025em b}\kern-.08em
    T\kern-.1667em\lower.7ex\hbox{E}\kern-.125emX}}
\DeclareMathOperator*{\argmax}{argmax}
\DeclareMathOperator*{\argmin}{argmin}
\DeclareMathOperator*{\dham}{d_H}
\DeclareMathOperator*{\sign}{sign}
\DeclareMathOperator*{\diag}{diag}
\DeclareMathOperator*{\R}{R}
\DeclareMathOperator*{\U}{U}
\DeclareMathOperator*{\p}{p}

\begin{document}

\title{Pseudo-Random Quantization Based Two-Stage Detection in One-Bit Massive MIMO Systems}

\author{Gökhan Yılmaz, \IEEEmembership{Graduate Student Member, IEEE}, and\\Ali Özgür Yılmaz, \IEEEmembership{Member, IEEE}
\thanks{The work of G. Yılmaz was supported by Vodafone Turkey within the framework of 5G and Beyond Joint Graduate Support Program coordinated by the Information and Communication Technologies Authority of Turkey.}
\thanks{G. Yılmaz is with the Department of Electrical Engineering, Eindhoven University of Technology, Eindhoven, The Netherlands (e-mail: g.yilmaz@tue.nl) and A. Ö. Yılmaz is with the Department of Electrical and Electronics Engineering, Middle East Technical University, Ankara, Turkey. (e-mail: aoyilmaz@metu.edu.tr)}
}

\maketitle

\begin{abstract}
Utilizing low-resolution analog-to-digital converters (ADCs) in uplink massive multiple-input multiple-output (MIMO) systems is a practical solution to decrease power consumption. The performance gap between the low and high-resolution systems is small at low signal-to-noise ratio (SNR) regimes. However, at high SNR and with high modulation orders, the achievable rate saturates after a finite SNR value due to the stochastic resonance (SR) phenomenon. This paper proposes a novel pseudo-random quantization (PRQ) scheme by modifying the quantization thresholds that can help compensate for the effects of SR and makes communication with high-order modulation schemes such as $1024$-QAM in one-bit quantized uplink massive MIMO systems possible. Moreover, modified linear detectors for non-zero threshold quantization are derived, and a two-stage uplink detector for single-carrier (SC) multi-user systems is proposed. The first stage is an iterative method called Boxed Newton Detector (BND) that utilizes Newton's Method to maximize the log-likelihood with box constraints. The second stage, Nearest Codeword Detector (NCD), exploits the first stage solution and creates a small set of most likely candidates based on sign constraints to increase detection performance. The proposed two-stage method with PRQ outperforms the state-of-the-art detectors from the literature with comparable complexity while supporting high-order modulation schemes.
\end{abstract}


\section{Introduction} \label{sec:intro}
\IEEEPARstart{M}{assive} multiple-input multiple-output (MIMO) systems can increase the spectral efficiency in multi-user settings by orders of magnitude due to much larger multiplexing and diversity potential \cite{Massive_MIMO_Survey}. However, since each antenna requires a separate radio-frequency (RF) chain, power consumption and hardware cost can increase to unacceptable levels. Analog-to-digital converter (ADC) units are one of the most power-consuming components in RF chains. The power consumption of an ADC increases linearly with the sampling rate and exponentially with the resolution. Therefore, low-resolution ADCs have become an attractive solution \cite{Low_Res_ADC_Survey}. Especially, one-bit ADCs have additional benefits, such as simple implementation and operation without automatic gain control (AGC) units. However, the received signal is severely distorted due to one-bit quantization.

The achievable rate \cite{CE_and_Perf_Analysis, Perf_Bound_Rayleigh_Channel, Throughput_Analysis, Wideband_1bit_Perf} and the channel capacity \cite{Comm_Limits_Low_Res_ADC, CSIT_Capacity, Capacity_High_SNR_mmWave} of one-bit systems are well-studied. The performance gap between the one-bit and infinite-resolution systems is small at low signal-to-noise ratio (SNR), yet it gets more prominent as SNR increases, which is undesirable in achieving higher rates per user. Stochastic resonance (SR) is observed when the presence of white noise in a nonlinear system's input increases the detectability of the input signal \cite{sr_detectability}. Different examples of systems where SR is encountered can be found in \cite{Stoch_Res_Review,sr_dynamic, Medical_Imaging_SR}. The intentional addition of noise into a system's input to reduce quantization errors is called dithering \cite{Stoch_Res_as_Dither}. Dithering can be achieved by adding artificially generated noise into the quantizer input, but it can also be achieved by just changing the quantization thresholds \cite{dafsp}. Non-zero threshold quantization has been utilized in some of the previous works in literature. In \cite{Deep_Signal_Recov}, deep unfolding is utilized to reconstruct a signal from one-bit measurements, with uniformly distributed thresholds. Threshold designs with a set partitioning scheme in \cite{Generalized_Bussgang_CE} and an adaptive technique in \cite{Quant_Design_and_CE} are utilized to minimize channel estimation errors. In \cite{Antithetic_Dither_CE}, an antithetic dithering scenario is exploited to increase the estimation performance. Changing the sampling characteristics can also be helpful to obtain better performance, examples of which can be found in \cite{dafsp, Perf_of_FTSR_Sampling, Oversampling_MIMO_OFDM}.

Linear detectors for one-bit massive MIMO systems have been studied extensively. They can be separated into two classes: conventional and Bussgang-based, as in \cite{ADMM, OBMNet}. The conventional linear detectors \cite{CE_and_Perf_Analysis, Throughput_Analysis} are quantization-unaware and constructed as if the system is infinite-resolution. Bussgang-based methods account for quantization and can offer better performance and lower error floors than their conventional counterparts. Although linear detectors are easy to implement, higher rates can be achieved with more sophisticated approaches. In \cite{Bayes_Optimal_CE_DD}, a Bayes-optimal joint channel estimation and detection scheme is proposed. \cite{ADMM} utilizes an ADMM-based algorithm for detection. A message passing algorithm for CP-free systems operating in ISI channels is proposed in \cite{Ungerboeck_Rec}. A recent study in \cite{Threshold_Rec_Design_and_Strategies} utilizes a hybrid scheme with analog processing and adaptive quantization thresholds to maximize the achievable rate.

In \cite{nML}, a near maximum likelihood (ML) approach is presented with a projected gradient descent algorithm, and \cite{1BOX} proposes a gradient descent algorithm with box constraints for one-bit MIMO systems that employ orthogonal frequency division multiplexing (OFDM). The works in \cite{Deep_Signal_Recov, OBMNet, LoRDNet} use gradient descent with the deep unfolding technique. Other detectors that rely on machine learning can be found in \cite{SVM, Robust_DD_Reinf_Learning, Supervised_Learning_Comm_Framework}. The detectors in \cite{nML, SVM, OBMNet} follow a two-stage approach. The second stage in \cite{nML} improves the performance by generating a reduced set to apply ML detection. The complexity is limited by a nearest neighbor search algorithm in \cite{OBMNet}. In \cite{SVM}, the final decisions are made using a metric called weighted Hamming distance \cite{Hamm_Dist_Decod}. 

In this paper, we focus on a new quantization strategy for one-bit massive MIMO systems that exploits pseudo-randomly generated non-zero quantization thresholds by generating a dithering effect to better recover the amplitude of received signal and to enhance detection performance. We first modify the linear filters to work with non-zero threshold quantization. Then, we focus on a new two-stage detection scheme. The first stage works as an equalizer to obtain an estimate of the signal to be detected by approximating ML detection to an unconstrained optimization problem for which Newton's method is utilized. Then, in the second stage, the first-stage estimate is used to create a set of most likely candidates to finalize decision-making. This two-stage procedure can help obtain an approximate-ML detection method that is computationally efficient.

Next, we introduce a new quantization strategy which is a reminiscent of the conventional dithering approach that relies on modifying the quantization thresholds of ADCs instead of generating analog dither signals. This strategy involves generating pseudo-random quantization thresholds and uses the threshold information during baseband processing to increase detection performance while not requiring additional circuitry. A critical advantage of this approach is not requiring temporal updates for different channel realizations and not requiring modification as long as the average SNR remains the same by exploiting the properties of massive MIMO. We analyze the performance of the PRQ scheme by both information theoretic and coding theoretic approaches to show that the proposed PRQ scheme can outperform the conventional zero-threshold quantization (ZTQ) scheme adopted by the previous detection studies in the literature \cite{1BOX,Bayes_Optimal_CE_DD,OBMNet,nML,Robust_DD_Reinf_Learning,Throughput_Analysis,SVM,Supervised_Learning_Comm_Framework,sc_fde,ADMM} without requiring additional power and processing. Also, the simulation results show that by combining the proposed detection and quantization schemes, one-bit massive MIMO systems can operate with high-order modulation schemes beyond QPSK or $16$-QAM in multi-user settings.

\subsection{Contributions}

We propose a novel PRQ scheme that can help mitigate the negative effects of SR. The proposed scheme relies on changing the quantization thresholds for dithering without requiring additional processing and circuitry. An achievable rate analysis and a minimum Hamming distance analysis are made to demonstrate the advantage of PRQ compared to the conventional ZTQ. Modified linear detectors for non-zero threshold quantization are derived. With the appropriate scaling factor, conventional linear receivers perform very closely to their Bussgang-based counterparts. The proposed first stage, named Boxed Newton Detector (BND), relies on Newton's Method with box constraints to generate an estimate of the input. It outperforms the existing gradient-based detectors \cite{nML, SVM, OBMNet} in terms of error performance with comparable complexity, while not requiring different hyperparameter selections for different system setups. With a similar motivation as in \cite{OBMNet}, the proposed second stage, named Nearest Codeword Detector (NCD), creates a set of candidate vectors to refine the first stage solution. However, unlike \cite{OBMNet}, it does so by taking the one-bit observations as binary codewords to find a limited number of candidates based on the minimum Hamming distance criterion to lower complexity and to increase detection performance. The proposed scheme can outperform the existing detectors in the literature with much lower error floors. Moreover, communication with high-order modulations such as $256$-QAM, $1024$-QAM, and $4096$-QAM, whose performances were not reported by any of the previous work in the literature, is shown to be possible with PRQ.

\subsection{Notation}

Lower-case letters represent scalars, lower-case bold letters represent column vectors, and upper-case bold letters represent matrices. $(.)^T$ is the transpose, and $(.)^H$ is the Hermitian. The $n^{\textrm{th}}$ element of a vector $\boldsymbol{c}$ is $c_n$. The $n^{\textrm{th}}$ row and $m^{\textrm{th}}$ column of a matrix $\boldsymbol{C}$ is denoted as $[\boldsymbol{C}]_{n,m}$. $\diag(.)$ creates a diagonal matrix. $\mathbb{E}[.]$ is the expectation operator. $|.|$ denotes the absolute value of its scalar arguments and the cardinality of its set arguments. $\mathbb{R}$ and $\mathbb{C}$ are the set of real and complex numbers, respectively. $\Re\{.\}$ and $\Im\{.\}$ give their arguments' real and imaginary parts, respectively. $\mathcal{CN}(\boldsymbol{\mu},\boldsymbol{\Sigma})$ denotes the complex Gaussian distribution with mean vector $\boldsymbol{\mu}$ and covariance matrix $\boldsymbol{\Sigma}$. $\phi(x)=\sqrt{\frac{1}{2\pi}} \exp(-\frac{x^2}{2})$ is the probability density function (PDF) of the standard Gaussian distribution.

\section{System Model} \label{sec:system_model}

\begin{figure}[b]
\centering
\includegraphics[width=\columnwidth*2
/5]{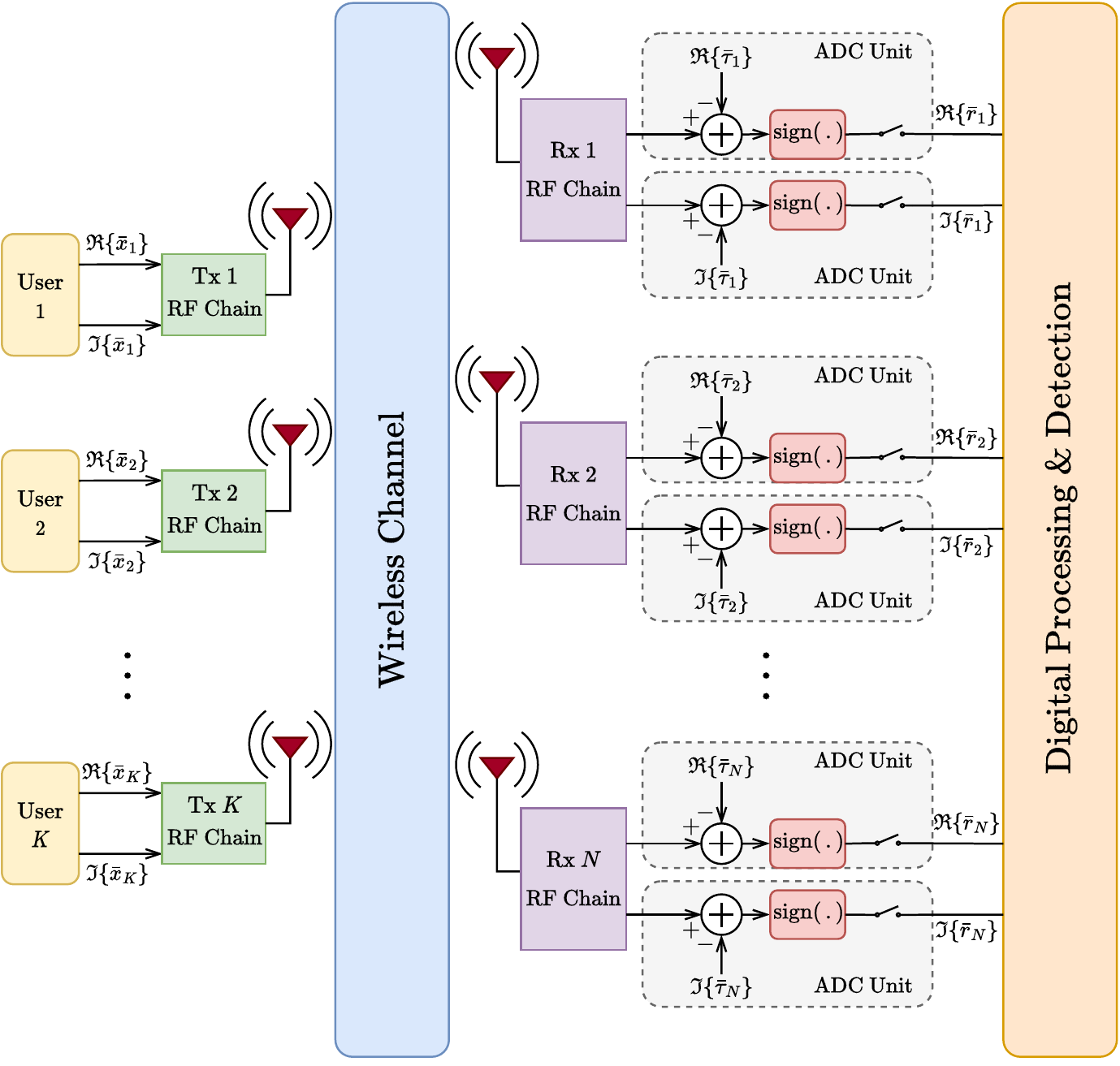}
\caption{A block diagram that summarizes the system model.}
\label{fig:system_model}
\end{figure}

We consider an uplink massive MIMO system where $K$ single-antenna users are served by a BS equipped with $N$ antennas. A simple block diagram of the system model is shown in Fig. \ref{fig:system_model}. Each user randomly selects an equally likely symbol from an $M$-QAM alphabet denoted by $\bar{\mathcal{M}}$. The vector of transmitted symbols is denoted as $\bar{\boldsymbol{x}}$, where $\bar{x}_k \in \bar{\mathcal{M}}$, $\mathbb{E}[\bar{x}_k]=0$, and $\mathbb{E}[|\bar{x}_k|^2]=E_s=1$ for $k=0, 1, \hdots, K$. Users transmit their signals through a Nyquist pulse-shaping filter, an ideal digital-to-analog converter (DAC), and I/Q modulation.

We assume that the channel impulse response (CIR) between each user and receiver antenna is perfectly known by the BS and can be modeled with uncorrelated Rayleigh fading. The BS performs decision-making in this uncoded uplink scenario to detect the symbols sent from all users by using the channel state information (CSI) knowledge and perfectly known quantization thresholds of ADCs in each RF chain. Additional analysis about the imperfect CSI scenario is also included in Section \ref{sec:num_results}. The CIR between receiver antenna $n$ and user $k$ has complex Gaussian distribution, i.e., $\bar{h}_{(n,k)} \sim \mathcal{CN}(0,1)$ as in many massive MIMO studies \cite{1BOX,ADMM,Wideband_1bit_Perf,Throughput_Analysis,CSIT_Capacity,LoRDNet,OBMNet,SVM,Supervised_Learning_Comm_Framework}. Hence, we investigate the effects of only small-scale fading. When pathloss and shadowing are taken into account, their effects can be considered as part of the SNR scaling. The equal average received signal power per user model can be realized using the advanced power control capabilities of the 5G networks. The resultant $N \times K$ channel matrix is represented as $[\bar{\boldsymbol{H}}]_{n,k}=\bar{h}_{(n,k)}$. Assuming perfect synchronization, after the received signal is I/Q demodulated, pulse matched-filtered, and symbol-rate sampled, the unquantized discrete-time signal is obtained as

\begin{equation}
\label{eqn:sc_unq_io_relation_complex}
\bar{\boldsymbol{y}} = \bar{\boldsymbol{H}} \bar{\boldsymbol{x}} + \bar{\boldsymbol{w}},
\end{equation}

\noindent where $\bar{\boldsymbol{w}} \sim \mathcal{CN}(\boldsymbol{0}_N,N_0 \boldsymbol{I}_N)$ is the zero-mean circularly symmetric Gaussian noise vector. $\boldsymbol{I}_N$ is the identity matrix of size $N \times N$ and $\boldsymbol{0}_N$ is the all-zero vector of size $N$. The received signal is quantized by a pair of one-bit ADCs at each receiver antenna. The unquantized samples and the quantization threshold vector $\bar{\boldsymbol{\tau}} = \Re\{ \bar{\boldsymbol{\tau}} \} + j \Im\{ \bar{\boldsymbol{\tau}} \}$ can be used to obtain the quantized signal

\begin{equation}
\label{eqn:sc_quant_io_relation_complex}
\bar{\boldsymbol{r}} = \sign(\Re\{\bar{\boldsymbol{y}}-\bar{\boldsymbol{\tau}}\}) + j \, \sign(\Im\{\bar{\boldsymbol{y}}-\bar{\boldsymbol{\tau}}\}),
\end{equation}

\noindent where $\sign(x)=+1$, if $x \geq 0$, and $\sign(x)=-1$ if $x < 0$. We define the SNR as the average SNR of each user at each receiver antenna:

\begin{equation}
\label{eqn:snr_def}
\rho = \frac{\mathbb{E}[|\bar{x}_k|^2]}{\mathbb{E}[|\bar{w}_n|^2]} = \frac{E_s}{N_0} = \frac{1}{N_0},
\end{equation}

\noindent since $\mathbb{E}[|h_{(n,k)}|^2] = 1$ for all $k=1,\hdots,K$ and $n=1,\hdots,N$.

The relation between a complex vector $\bar{\boldsymbol{a}}$ and its real counterpart $\boldsymbol{a}$, and a complex matrix $\bar{\boldsymbol{A}}$ and its real counterpart $\boldsymbol{A}$ can be obtained as

\begin{equation}
    \label{bnd_eqn:vector_real}
    \boldsymbol{a} =
    \begin{bmatrix}
    \Re\{\bar{\boldsymbol{a}}\} \\ \Im\{\bar{\boldsymbol{a}}\}
    \end{bmatrix} \textrm{ and }
    \boldsymbol{A} =
    \begin{bmatrix}
    \Re\{\bar{\boldsymbol{A}}\} & -\Im\{\bar{\boldsymbol{A}}\} \\
    \Im\{\bar{\boldsymbol{A}}\} & \Re\{\bar{\boldsymbol{A}}\}
    \end{bmatrix}.
\end{equation}

As a result, the overall input-output relation of the system can also be expressed as

\begin{equation}
\label{eqn:sc_quant_io_relation_real}
\boldsymbol{r} = \sign(\boldsymbol{H} \boldsymbol{x} + \boldsymbol{w}- \boldsymbol{\tau}).
\end{equation}

Note that each element of $\boldsymbol{x}$ belongs to the set of values a symbol from the modulation alphabet can take in one dimension. We denote this set as $\mathcal{M}$. For example, when $\bar{\mathcal{M}}$ is the $16$-QAM alphabet, $\mathcal{M}$ can be expressed as the $4$-PAM alphabet with an average power of $1/2$.

\section{First Stage Linear Detectors} \label{sec:linear_detectors}

\subsection{Bussgang-Based Linear Filters}

The Bussgang decomposition \cite{Bussgang_Dec} is valid only when the input to the quantizer has Gaussian distribution. Despite $\bar{\boldsymbol{y}}$ not having Gaussian distribution, especially at low SNR when the noise term is dominant or due to the central limit theorem (CLT) when the number of users is high, the distribution of $\bar{\boldsymbol{y}}$ is very close to Gaussian \cite{CE_and_Perf_Analysis, Throughput_Analysis}. Therefore, applying the Bussgang decomposition for designing and analyzing low-resolution systems is a common practice \cite{Bussgang_Dec, CE_and_Perf_Analysis, Threshold_Rec_Design_and_Strategies, Throughput_Analysis, Oversampling_MIMO, ADMM, OBMNet}. Since we focus on a more general scenario where thresholds can take arbitrary values, an extended version of the Bussgang theorem must be used. In \cite{Generalized_Bussgang_CE}, the Bussgang theorem is generalized for the non-zero threshold quantization scenario by taking the selection process of the Bussgang gain and bias terms as the problem of finding the minimum mean square error (MMSE) estimate of the quantized observation vector $\Bar{\boldsymbol{r}}$ using the unquantized observation $\Bar{\boldsymbol{y}}$ such that

\begin{equation}
    \Bar{\boldsymbol{r}} - \Bar{\boldsymbol{b}} = \Bar{\boldsymbol{g}} \, \Bar{\odot} \, \Bar{\boldsymbol{y}} + \Bar{\boldsymbol{d}}, 
\end{equation}

\noindent where $\Bar{\boldsymbol{b}} = \mathbb{E}[\Bar{\boldsymbol{r}} \mid \Bar{\boldsymbol{\tau}}]$ is the bias vector, $\Bar{\boldsymbol{g}} \in \mathbb{C}^{N}$ is the Bussgang gain vector which can also be represented as a diagonal matrix composed of the entries of this vector, i.e., the MMSE filter, and $\Bar{\boldsymbol{d}} \in \mathbb{C}^{N}$ is the quantization noise vector. Note that unlike \cite{Generalized_Bussgang_CE}, we adopt a notation with complex numbers by defining a complex extension to the Hadamard product such that

\begin{equation}
    \label{eqn:complex_odot}
    \bar{\boldsymbol{u}} \, \bar{\odot} \, \bar{\boldsymbol{v}} = \Re\{\bar{\boldsymbol{u}}\} \odot \Re\{\bar{\boldsymbol{v}}\} + j \Im\{\bar{\boldsymbol{u}}\} \odot \Im\{\bar{\boldsymbol{v}}\}.
\end{equation}

Using the derivations from \cite{Generalized_Bussgang_CE}, the Bussgang gain vector and the bias vector can be calculated respectively as

\begin{equation}
    \label{eqn:buss_gain_complex}
    \bar{g}_n = \sqrt{\frac{4}{\pi [\bar{\boldsymbol{C}}_y]_{(n,n)}}} \left( \exp \left( - \frac{\Re\{\bar{\tau}_n\}^2}{[\bar{\boldsymbol{C}}_y]_{(n,n)}} \right) + j \exp \left( - \frac{\Im\{\bar{\tau}_n\}^2}{[\bar{\boldsymbol{C}}_y]_{(n,n)}} \right) \right),
\end{equation}

\begin{equation}
    \label{eqn:bias_complex}
    \bar{b}_n = 2 \Phi \left( - \frac{ \Re\{\bar{\tau}_n\} }{ \sqrt{ [{\bar{\boldsymbol{C}}}_y]_{(n,n)}/2 } } \right) - 1 + j \left( 2 \Phi \left( - \frac{ \Im\{\bar{\tau}_n\} }{ \sqrt{ [\bar{\boldsymbol{C}}_y]_{(n,n)}/2 } } \right) - 1 \right),
\end{equation}

\noindent where $\Bar{\boldsymbol{C}}_y = \mathbb{E}[\Bar{\boldsymbol{y}} \Bar{\boldsymbol{y}}^H] = \Bar{\boldsymbol{H}} \Bar{\boldsymbol{H}}^H + N_0 \boldsymbol{I}_N$ is the covariance matrix of the unquantized observation vector. For ease of notation, we define the conditionally zero-mean version of the quantized observation vector such that

\begin{equation}
    \label{eqn:bussgang_io_flat}
        \bar{\boldsymbol{r}}_e =
        \bar{\boldsymbol{r}} - \bar{\boldsymbol{b}} = \bar{\boldsymbol{g}} \, \bar{\odot} \, \bar{\boldsymbol{H}} \bar{\boldsymbol{x}} + \bar{\boldsymbol{g}} \, \bar{\odot} \, \bar{\boldsymbol{w}} + \bar{\boldsymbol{d}} \\
        = \bar{\boldsymbol{H}}_e \bar{\boldsymbol{x}} + \bar{\boldsymbol{w}}_e,
\end{equation}

\noindent where $\bar{\boldsymbol{H}}_e \in \mathbb{C}^{N \times K}$ is the effective channel matrix which takes the Bussgang gains into account and $\bar{\boldsymbol{w}}_e \in \mathbb{C}^{N}$ is the effective noise vector which is the combination of quantized thermal noise and quantization distortion. Now that we defined the linearized version of the input-output relation, we can move on to define the Bussgang-based MRC (BMRC) and ZF (BZF) filters as

\begin{equation}
    \label{nhd:bmrc_filter}
    \bar{\boldsymbol{F}}_{\mathrm{BMRC}} = \bar{\boldsymbol{\lambda}}^e \odot \bar{\boldsymbol{H}}_e^H \textrm{ and }     \bar{\boldsymbol{F}}_{\mathrm{BZF}} = (\bar{\boldsymbol{H}}_e^H \bar{\boldsymbol{H}}_e)^{-1} \bar{\boldsymbol{H}}_e^H,
\end{equation}

\noindent where $\bar{\boldsymbol{\lambda}}^e$ is the scaling constant to make sure we obtain a non-scaled estimate, and it is determined as $
    \bar{\lambda}_k^e = \sfrac{1}{\sum_{n=1}^{N} |[\bar{\boldsymbol{H}}_e]_{(n,k)}|^2}$, for $k=1,2,\hdots,K$.

Once a filter $\Bar{\boldsymbol{F}}$ to be applied on the conditionally zero-mean observation vector $\Bar{\boldsymbol{r}}_e$ is selected, the estimates can be obtained as

\begin{equation}
    \label{eqn:filtering}
    \Tilde{\Bar{\boldsymbol{x}}} = \Bar{\boldsymbol{F}} \Bar{\boldsymbol{r}}_e.
\end{equation}

Finally, if no further processing is to be applied, decisions can be obtained by symbol-by-symbol detection, i.e., element-wise minimum distance mapping.

\subsection{Conventional Linear Filters} \label{subsec:conv_lin}

Since one-bit measurement causes the loss of amplitude information, inserting appropriate labels to our quantized observations is a helpful way to work with variable-amplitude modulation schemes. To do so, we continue with the Bussgang decomposition and look for approximations that make all entries of $\Bar{\boldsymbol{g}}_n$ the same in both the I and Q parts. To do so, we first assume the threshold values are unknown and $\Bar{\boldsymbol{\tau}} \sim \mathcal{CN}(\boldsymbol{0}_N, \sigma^2_{\tau})$. This can be a useful assumption if the empirical distribution of the thresholds is close to Gaussian. Then randomly selected thresholds can be seen as an additional form of noise over additive white Gaussian noise (AWGN). Therefore the effective noise variance becomes $\sigma^2_e = N_0 + \sigma^2_{\tau}$ since the thermal noise and threshold selections are independent. With this selection, we can also act as if $\Bar{\boldsymbol{\tau}} = \boldsymbol{0}_N$. Now, the effective unquantized observation covariance matrix becomes $\Bar{\boldsymbol{C}}_{y,e} = \Bar{\boldsymbol{H}} \Bar{\boldsymbol{H}}^H + (N_0 + \sigma^2_{\tau}) \boldsymbol{I}_N$.

Finally, the wireless channel's effect must be considered to obtain a label that is the same for all branches. At low SNR or when the number of users is large, the unquantized observation covariance matrix can be approximated by a diagonal matrix such that $\Bar{\boldsymbol{C}}_{y,e} \cong (K+N_0+\sigma^2_{\tau}) \boldsymbol{I}_N$. By using these approximations, the bias vector becomes $\Bar{\boldsymbol{b}} = \boldsymbol{0}_N$ and each element of the Bussgang gain vector can be approximated as $\Bar{g}_n \cong \sqrt{\frac{4}{\pi (K+N_0+\sigma^2_{\tau})}} (1 + j)$. Since the I and Q parts are now the same, the input-output relation can be simplified as

\begin{equation}
    \label{eqn:simplified_bussgang}
    \Bar{\boldsymbol{r}}_e = \Bar{\boldsymbol{r}} = \sqrt{\frac{4}{\pi (K+N_0+\sigma^2_{\tau})}} \Bar{\boldsymbol{y}} + \Bar{\boldsymbol{d}}.
\end{equation}

By defining the quantization label as $\ell_q = \sqrt{\sfrac{\pi (K+N_0+\sigma_{\tau}^2)}{4}}$ and multiplying both sides of (\ref{eqn:simplified_bussgang}) with $\ell_q$, we can get

\begin{equation}
    \label{eqn:simplified_bussgang2}
    \ell_q \Bar{\boldsymbol{r}} = \Bar{\boldsymbol{y}} + \ell_q \Bar{\boldsymbol{d}}
    = \Bar{\boldsymbol{H}} \Bar{\boldsymbol{x}} + \Bar{\boldsymbol{w}} + \ell_q \Bar{\boldsymbol{d}}.
\end{equation}

Using this quantization label, a non-scaled estimate of the $\Bar{\boldsymbol{x}}$ vector can now be obtained when conventional linear filters are used. Finding the conventional MRC and ZF filters is now straightforward. The MRC and ZF filters can be calculated as

\begin{equation}
    \label{eqn:mrc}
    \bar{\boldsymbol{F}}_{\mathrm{MRC}} = \bar{\boldsymbol{\lambda}} \odot \bar{\boldsymbol{H}}^H \textrm{ and } \bar{\boldsymbol{F}}_{\mathrm{ZF}} = (\bar{\boldsymbol{H}}^H \bar{\boldsymbol{H}})^{-1} \bar{\boldsymbol{H}}^H,
\end{equation}

\noindent where the elements of scaling vector $\Bar{\boldsymbol{\lambda}}$ for $k=1,2,\hdots,K$ can be found as $
    \bar{\lambda}_k[v] = \sfrac{1}{\sum_{n=1}^{N} |[\bar{\boldsymbol{H}}_{(n,k)}|^2}$.

Once a filter of choice $\Bar{\boldsymbol{F}}$ is selected, it is applied on the quantized observation vector $\Bar{\boldsymbol{r}}$ to obtain the estimates as

\begin{equation}
    \label{eqn:filtering2}
    \Tilde{\Bar{\boldsymbol{x}}} = \ell_q \Bar{\boldsymbol{F}} \Bar{\boldsymbol{r}}.
\end{equation}

Lastly, if no further processing is applied, decisions can be obtained using element-wise minimum distance mapping.

\section{Maximum Likelihood (ML) Detector} \label{sec:ml_detector}

Likelihood-based detectors have an important advantage against linear detectors in terms of bit error rate (BER) performance, especially at high SNR \cite{OBMNet}. As explained in different works such as \cite{nML,Deep_Signal_Recov,OBMNet}, the conditional probability mass function (PMF) of the quantized observations can be expressed as

\begin{equation}
\label{eqn:vector_cond_pmf}
\p(\boldsymbol{r} | \boldsymbol{x}, \boldsymbol{\tau}, \boldsymbol{H})
= \prod_{n=1}^{2N} \Phi \left( \frac{r_n (\boldsymbol{h}_n^T \boldsymbol{x} - \tau_n)}{\sqrt{N_0/2}} \right),
\end{equation}

\noindent where $\boldsymbol{h}_n^T$ is the $n^{\textrm{th}}$ row of the channel matrix $\boldsymbol{H}$, i.e., $\boldsymbol{H}=\begin{bmatrix} \boldsymbol{h}_1 & \boldsymbol{h}_2 & \hdots & \boldsymbol{h}_{2N} \end{bmatrix}^T$, and $\Phi(\alpha) = \int_{-\infty}^{\alpha} \phi(\tau) \, d\tau$ is the cumulative distribution function (CDF) of the standard Gaussian random variable. Then, by utilizing the log-likelihood, we can construct the ML detector as

\begin{equation}
\label{eqn:ml_detector_long}
\hat{\boldsymbol{x}}_{\textrm{ML}}
= \argmax_{\boldsymbol{x} \in \mathcal{M}^{2K}} \left\{ \boldsymbol{1}_N \ln \left( \Phi \left( \sqrt{\frac{2}{N_0}} \boldsymbol{r} \odot (\boldsymbol{H} \boldsymbol{x} - \boldsymbol{\tau}) \right) \right) \right\},
\end{equation}

\noindent where the natural logarithm $\ln(.)$ and $\Phi(.)$ are applied element-wise on their arguments.

\section{Proposed First Stage: Boxed Newton Detector (BND)} \label{sec:BND}

When the modulation order $M$ or the number of users $K$ is large, complexity becomes infeasible for ML detection. Different works in the literature focus on a gradient-based approach to maximize the log-likelihood function \cite{nML, Deep_Signal_Recov, OBMNet, LoRDNet}. In this section, a similar procedure is followed with a new iterative receiver that uses a modified cost function and the Hessian information for faster convergence via Newton's method. For compactness, we denote the log-likelihood as

\begin{equation}
    \mathcal{L}(\boldsymbol{x}) = \boldsymbol{1}_N \ln \left( \Phi \left( \sqrt{\frac{2}{N_0}} \boldsymbol{r} \odot (\boldsymbol{H} \boldsymbol{x} - \boldsymbol{\tau}) \right) \right),
\end{equation}

\noindent where $\boldsymbol{1}_N$ is a vector of size $N$ whose elements are all set to $1$. We approximate the ML detection problem as a new optimization problem where we relax the constraint that $\boldsymbol{x}$ belongs to a discrete input set and add a penalty function to reflect the box constraints related to the boundary of the chosen constellation scheme to compensate for the relaxation and obtain favorable convergence. The penalty method is useful for keeping the formulation as an unconstrained optimization problem while using the constellation boundaries as a priori information. The boundary of an $M$-QAM type constellation with unit average power on a single dimension can be calculated as

\begin{equation}
\label{eqn:constraint}
M_b = \sqrt{\frac{3 (\sqrt{M}-1)^2}{2 (M-1)}}.
\end{equation}

The new unconstrained optimization problem formulation with the penalty function $\mathcal{P}(.)$ can be expressed as

\begin{equation}
\label{eqn:unconstrained_opt}
\tilde{\boldsymbol{x}}
= \argmax_{\boldsymbol{x} \in \mathbb{R}^{2K}} \{ \mathcal{L}(\boldsymbol{x}) - \mathcal{P}(\boldsymbol{x}) \},
\end{equation}

\noindent where the penalty function reflects how much deviation occurs from the boundaries of the constellation scheme which can be expressed as

\begin{equation}
\label{eqn:penalty_func}
\mathcal{P}(\boldsymbol{x})
= \frac{\theta}{2} \sum_{k=1}^{2K} \R(|x_k| - M_b)^2,
\end{equation}

\noindent where $\theta \in \mathbb{R}$ is a constant that determines the strength of box constraints, and $\R(x) = \max\{0,x\}$ is the unit-ramp function. Gradient-based optimization can be utilized for (\ref{eqn:unconstrained_opt}) due to the log-concavity of $\Phi(.)$ and the negative of the penalty function being concave as well. By denoting the argument of (\ref{eqn:unconstrained_opt}) as our cost function $\mathcal{J}(\boldsymbol{x}) = \mathcal{L}(\boldsymbol{x}) - \mathcal{P}(\boldsymbol{x})$, the iterative update equation of Newton's Method becomes

\begin{equation}
\label{eqn:newtons_update}
\boldsymbol{x}^{k+1}
= \boldsymbol{x}^{k} - (\nabla^2 \mathcal{J}(\boldsymbol{x}^{k}))^{-1} \nabla \mathcal{J}(\boldsymbol{x}^{k}),
\end{equation}

\noindent where $\nabla = \begin{bmatrix} \frac{\partial}{\partial x_1} & \hdots & \frac{\partial}{\partial x_{2K}} \end{bmatrix}^T$ can be expressed as a vector of size $2K$, and $\nabla ^2$ can be expressed as the outer product of the $\nabla$ operator with itself: $\nabla ^2 = \nabla \, \nabla ^T$. Hence, the $\nabla ^2$ operator can be considered a $2K \times 2K$ matrix. $\nabla \mathcal{J}(\boldsymbol{x}^{k})$ is the gradient of the cost function $\mathcal{J}$ with respect to $\boldsymbol{x}$ calculated at $\boldsymbol{x}^k$, i.e., the estimate of $\boldsymbol{x}$ at iteration $k$. Similarly, $\nabla ^2 \mathcal{J}(\boldsymbol{x}^{k})$ is the Hessian of the cost function $\mathcal{J}$ with respect to $\boldsymbol{x}$ calculated at $\boldsymbol{x}^k$.

Utilizing a second-order optimization technique such as Newton's method for this problem helps for getting fast convergence and avoids the burden of selecting different step-sizes for different system setups as opposed to \cite{nML, 1BOX}. Machine learning tools have been used to avoid this issue \cite{OBMNet, LoRDNet}. However, these methods require an offline learning stage and larger number of iterations compared to Newton's method, making these approaches computationally comparable. We should go over the log-likelihood and penalty functions' first and second-order derivatives before expressing the gradient and Hessian information. The first-order derivatives of the log-likelihood function and the penalty function can be expressed as

\begin{equation}
    \label{eqn:firstorder_lld}
    \frac{d}{dx_k} \mathcal{L}(\boldsymbol{x}) =  \sqrt{\frac{2}{N_0}} \sum_{n=1}^{2N} r_n [\boldsymbol{H}]_{(n,k)} \varphi \left( u_n \right),
\end{equation}

\begin{equation}
    \label{eqn:firstorder_pen}
    \frac{d}{dx_k} \mathcal{P}(\boldsymbol{x}) =  \theta \sign(x_k) \R(|x_k|-M_b),
\end{equation}

\noindent respectively, for $k=1,2,\hdots,2K$, where $\boldsymbol{u} = \sqrt{\frac{2}{N_0}} \boldsymbol{r} \odot (\boldsymbol{H} \boldsymbol{x} - \boldsymbol{\tau})$, and $\varphi(x) = \frac{d}{dx} \ln(\Phi(x))=\frac{\phi(x)}{\Phi(x)}$. Then the second-order derivatives can be obtained as

\begin{equation}
    \label{eqn:secorder_lld}
    \frac{d^2}{dx_k dx_m} \mathcal{L}(\boldsymbol{x}) = \frac{2}{N_0} \sum_{n=1}^{2N} [\boldsymbol{H}]_{(n,k)} [\boldsymbol{H}]_{(n,m)} \psi \left( u_n \right),
\end{equation}

\begin{equation}
    \label{eqn:secorder_pen}
    \frac{d^2}{dx_k dx_m} \mathcal{P}(\boldsymbol{x}) =  \theta \U(|x_k|-M_b) \delta[k-m],
\end{equation}

\noindent for $k=1,2,\hdots,2K$ and $m=1,2,\hdots,2K$, where $\psi(x) = \frac{d^2}{dx^2} \ln(\Phi(x)) = -x\varphi(x) -\varphi^2(x)$, $\U(x) = \max\{0,\sign(x)\}$ is the unit-step function, and $\delta[.]$ is the discrete unit impulse function.

Due to linearity of differentiation, we have $\nabla \mathcal{J}(\boldsymbol{x}) = \nabla \mathcal{L}(\boldsymbol{x}) - \nabla \mathcal{P}(\boldsymbol{x})$. As in \cite{Deep_Signal_Recov}, the gradient of the log-likelihood function can be found as

\begin{equation}
\label{eqn:gradient_L}
\nabla \mathcal{L}(\boldsymbol{x})
= \sqrt{\frac{2}{N_0}} \boldsymbol{H}^{T} \left(\boldsymbol{r} \odot \varphi \left( \boldsymbol{u} \right)\right),
 \end{equation}

\noindent where $\odot$ denotes the Hadamard product, and $\varphi(.)$ is applied element-wise on its arguments. Then, the gradient of the penalty function can be found as

\begin{equation}
\label{eqn:gradient_P}
\nabla \mathcal{P}(\boldsymbol{x})
= \theta \sign(\boldsymbol{x}) \odot \R(|\boldsymbol{x}| - M_b \boldsymbol{1}_{2K}),
 \end{equation}

\noindent where $\R(.)$ is applied element-wise on its arguments. Similarly, we have $\nabla ^2 \mathcal{J}(x) = \nabla ^2 \mathcal{L}(x) - \nabla ^2 \mathcal{P}(x)$. The Hessian of the log-likelihood function can be calculated as

\begin{equation}
\label{eqn:hessian_L}
\nabla^2 \mathcal{L}(\boldsymbol{x})
= \frac{2}{N_0} \boldsymbol{H}^{T} \diag \left( \psi \left( \boldsymbol{u}\right) \right) \boldsymbol{H},
\end{equation}

\noindent where $\psi(.)$ is applied element-wise on its arguments. Since $\Phi(x)$ can approach zero exponentially fast, computations of $\ln(\Phi(x))$, $\varphi(x)$, and $\psi(x)$ in finite precision can cause problems such as divergent behavior or uncertainties. This problem is solved in Appendix \ref{app:nonlinear_func}. Finally, the Hessian of the penalty function is

\begin{equation}
\label{eqn:hessian_P}
\nabla^2 \mathcal{P}(\boldsymbol{x})
= \theta \diag(\U(|\boldsymbol{x}| - M_b \boldsymbol{1}_{2K})),
\end{equation}

\noindent where $\U(.)$ is applied element-wise on its arguments.

\begin{algorithm}[tb!]
\scriptsize
\caption{Boxed Newton Detector (BND)}
\begin{algorithmic}[1]
\label{alg:bnd}
\renewcommand{\algorithmicrequire}{\textbf{Input:}}
\renewcommand{\algorithmicensure}{\textbf{Output:}}
\REQUIRE $\boldsymbol{r}, \boldsymbol{H}, \boldsymbol{\tau}, \theta, \epsilon, M_b, T_{\textrm{max}}, N_0, \zeta$
\ENSURE  $\tilde{\boldsymbol{x}}$
\STATE Set the initial solution to the MRC estimate $\tilde{\boldsymbol{x}} \gets \boldsymbol{x}^0$ using (\ref{eqn:mrc}) \label{line:initialization}
\STATE Apply the damping factor $N_0 \gets \zeta N_0$ \label{line:damping}
\FOR {$i = 1$ to $T_{\textrm{max}}$}
\STATE Calculate the gradient $\nabla \mathcal{J}(\tilde{\boldsymbol{x}})$ using (\ref{eqn:gradient_L}) and (\ref{eqn:gradient_P}) \label{line:gradient}
\STATE Calculate the Hessian $\nabla^2 \mathcal{J}(\tilde{\boldsymbol{x}})$ using (\ref{eqn:hessian_L}) and (\ref{eqn:hessian_P}) \label{line:hessian}
\STATE Calculate the step $\Delta \boldsymbol{x} \gets (\nabla^2 J(\tilde{\boldsymbol{x}}))^{-1} \nabla J(\tilde{\boldsymbol{x}})$ \label{line:step}
\STATE Iterative update $\tilde{\boldsymbol{x}} \gets \tilde{\boldsymbol{x}} - \Delta \boldsymbol{x}$ as in (\ref{eqn:newtons_update}) \label{line:update}
\IF {$\lVert \Delta \boldsymbol{x} \rVert ^2 < \epsilon \, \lVert \boldsymbol{x} \rVert ^2$} \label{line:termination}
\RETURN $\tilde{\boldsymbol{x}}$
\ENDIF
\ENDFOR
\RETURN $\tilde{\boldsymbol{x}}$
\end{algorithmic}
\end{algorithm}

Now that the update rule is wholly defined, an initial solution $\boldsymbol{x}^0$ that is preferably not far away from the final solution should be found. The MRC estimate is a suitable selection for this purpose, and it can easily be found using (\ref{eqn:mrc}). Note that at high SNR, when $N_0$ is very small, the Hessian matrix can become very close to singular since $\psi(x) \rightarrow 0$ as $x \rightarrow \infty$. To avoid such behavior, we define a damping factor $\zeta$ such that

\begin{equation}
\label{eqn:damping_factor}
\zeta = \max\{1,\rho/\rho_d\},
\end{equation}

\noindent where $\rho_d$ is the damping SNR. Before starting the iterative updates, $N_0$ is multiplied with this term to provide numeric stability. The complete procedure for BND is summarized in Algorithm \ref{alg:bnd}. Once the iterative updates start, the algorithm is terminated if the maximum number of iterations $T_{\mathrm{max}}$ is reached or if further iterations do not cause significant changes, which is determined by the termination threshold $\epsilon$. The output of the algorithm $\tilde{\boldsymbol{x}}$ is the first-stage solution, and if a one-stage approach is to be followed, then symbol-by-symbol detection is applied on the estimate. If not, the estimate is supplied to the second stage for further processing.

\section{Proposed Second Stage: Nearest Codeword Detector (NCD)} \label{sec:NCD}

After finding an estimate using a first-stage method, NCD can be utilized to make more accurate decisions compared to symbol-by-symbol detection. In this part, the first step is to decide on the reliability of each element of the first-stage estimate. Then, a set of candidate vectors is formed based on the reliability information of each element, very similar to the ideas from \cite{nML, SVM, OBMNet}. The candidate set is then narrowed down based on the minimum Hamming distance criterion in the codeword domain. Finally, ML detection is conducted on the smaller candidate set to make the final decisions. The complete summary of the proposed second stage detector is made in Algorithm \ref{alg:NCD}, and the detailed explanation of the whole procedure is made in the following parts.

\begin{algorithm}[tb]
\scriptsize
\caption{Nearest Codeword Detector (NCD)}
\begin{algorithmic}[1]
\label{alg:NCD}
\renewcommand{\algorithmicrequire}{\textbf{Input:}}
\renewcommand{\algorithmicensure}{\textbf{Output:}}
\REQUIRE $\tilde{\boldsymbol{x}}, \boldsymbol{r}, \boldsymbol{H}, \boldsymbol{\tau}, P, \gamma, U_{\mathrm{max}}$
\ENSURE  $\hat{\boldsymbol{x}}$
\STATE Find the nearest decision boundaries $\boldsymbol{t}$ as in (\ref{eqn:decision_boundaries}) \label{line:dec_bound}
\STATE Obtain the sets of reliable indices $\mathcal{R}$ as in (\ref{eqn:reliable_indices}) and unreliable indices $\mathcal{U}$ as in (\ref{eqn:unreliable_indices}) \label{line:rel_unrel_indices}
\WHILE{$|\mathcal{U}| > U_{\mathrm{max}}$} \label{line:while_start}
\STATE Decrease the size of the unreliable region as $\gamma \gets 0.95 \gamma$
\STATE Reobtain the sets $\mathcal{R}$ and $\mathcal{U}$ according to the new $\gamma$
\ENDWHILE \label{line:while_end}
\STATE Find the candidate element sets $(\mathcal{X}_k)_{k=1}^{2K}$ using (\ref{eqn:element_candidates})
\STATE Generate the candidate vector set $\mathcal{X}$ using (\ref{eqn:vector_candidates})
\IF{$|\mathcal{X}| > P$}
\STATE Apply symbol-by-symbol detection to get \\ $\check{x}_k \gets \argmin_{x \in \mathcal{M}} |\tilde{x}_k - x|$ for $k=1, 2, \hdots, 2K$
\STATE Remove the symbol-by-symbol detected vector from the set $\mathcal{X} \gets \mathcal{X} \setminus \{\check{\boldsymbol{x}}\}$
\IF{$|\mathcal{X}| > 1$}
\STATE Sort $\mathcal{X}$ according to the $\mathscr{H}(.)$ metric as in (\ref{eqn:candidate_sequence})
\STATE Discard all elements of $\mathcal{X}$ except the first $P-1$
\ENDIF
\STATE Add $\check{\boldsymbol{x}}$ back to the set to obtain the result from (\ref{eqn:final_set}) $\mathcal{X} \gets \mathcal{X} \cup \{\check{\boldsymbol{x}}\}$
\ENDIF
\STATE Apply ML detection on the set $\mathcal{X}$
\RETURN $\tilde{\boldsymbol{x}}$
\end{algorithmic} 
\end{algorithm}

The set of decision boundaries utilized during symbol-by-symbol detection for $M$-QAM constellations is found as

\begin{equation}
\label{eqn:decision_boundary_set}
\mathcal{T} = \left\{ \pm n \sqrt{\frac{6}{M-1}} \, \bigg| \, n \in \left\{0, 1, \hdots, \frac{\sqrt{M}-2}{2}\right\} \right\}.
\end{equation}

For example, the decision boundary set for QPSK constellation is $\{0\}$, and for $16$-QAM constellation, it is $\left\{0,\pm 2\sqrt{\frac{1}{10}}\right\}$. Then, the closest decision boundary to the $k^{\textrm{th}}$ element of the estimate $\tilde{\boldsymbol{x}}$ is defined as

\begin{equation}
\label{eqn:decision_boundaries}
t_k = \argmin_{t \in \mathcal{T}} |\tilde{x}_k - t|,
\end{equation}

\noindent for $k=1, 2, \hdots, 2K$. Then, the set of reliable and unreliable indices can be found as

\begin{equation}
\mathcal{R} = \{k \mid k \in \{1, 2, \hdots, 2K\}, \, |\tilde{x}_k - t_k| > \gamma \}, \label{eqn:reliable_indices}
\end{equation}

\begin{equation}
\mathcal{U} = \{k \mid k \in \{1, 2, \hdots, 2K\} \setminus \mathcal{R}\}, \label{eqn:unreliable_indices}
\end{equation}

\noindent respectively. If the $k^{\textrm{th}}$ element of the estimate is within a certain margin of its closest decision boundary defined by the hyperparameter $\gamma \in \mathbb{R}$, then it is not reliable to conduct symbol-by-symbol detection on $\tilde{x}_k$, for $k=1, 2, \hdots, 2K$. Now, the set of possible assignments for each element of the estimate can be found as

\begin{equation}
\label{eqn:element_candidates}
\tilde{\mathcal{X}}_k = \left\{
\begin{array}{ll}
\left\{\argmin_{x \in \mathcal{M}} |x_k -x| \right\}, & \quad k \in \mathcal{R} \\ \\
\left\{t_k \pm \sqrt{\frac{3}{2(M-1)}}\right\}, & \quad k \in \mathcal{U}
\end{array}
\right.
\end{equation}

\noindent for $k=1, 2, \hdots, 2K$. If the $k^{\textrm{th}}$ element of the estimate is reliable, then there is only one possible assignment to this element obtained by the minimum distance rule. If it is unreliable, the possible assignment is a set composed of two elements that are the neighbors to the closest decision boundary. Hence, each set's cardinality can be $1$ or $2$. The resultant set of candidate vectors from the combinations of each candidate element set can be obtained as

\begin{equation}
\label{eqn:vector_candidates}
\tilde{\mathcal{X}} = \tilde{\mathcal{X}}_1 \times \tilde{\mathcal{X}}_2 \times \hdots \times \tilde{\mathcal{X}}_{2K},
\end{equation}

\noindent where $\times$ denotes the Cartesian product operator. Examples of the set formation procedures shown in (\ref{eqn:element_candidates}) and (\ref{eqn:vector_candidates}) can be found in \cite{OBMNet}.

The cardinality of the candidate vector set $|\tilde{\mathcal{X}}|$ is at most $2^{2K}$, which means the set can grow exponentially as the number of users increases, which can cause two problems. First, the size of the set may get too large when $|\mathcal{U}|$ is large, which can cause memory problems and an undesirable complexity increase for ML detection at the final step. The first problem is dealt with adaptively changing $\gamma$, which can adjust the size of the unreliable region. If the size of $\mathcal{U}$ is large, i.e., it is greater than $U_{\mathrm{max}}$, then decreasing $\gamma$ can help us obtain a smaller unreliable region, hence a smaller set size for $\tilde{\mathcal{X}}$. For the second problem, similar to the idea from \cite{OBMNet}, the aim is to limit the search complexity by finding a subset $\mathcal{X}$ of $\tilde{\mathcal{X}}$ that includes the most likely candidates. In \cite{nML} and \cite{SVM}, the search complexity is not limited in the second stage, and in \cite{OBMNet}, a limited number of nearest neighbors to the estimate vector from the candidate set $\tilde{\mathcal{X}}$ are found. However, especially at high SNR, detection performance can benefit from the sign constraints imposed by one-bit quantization. Since quantized observations take binary values, $\boldsymbol{r}$ can easily be seen as a codeword in the spatial domain. Hence, the size of the set of candidate vectors can be limited by finding their codeword representations, then ordering them according to their Hamming distance to the actual quantized observation vector. The spatial codeword representation $c(.)$ of a candidate vector $\acute{\boldsymbol{x}} \in \tilde{\mathcal{X}}$ is defined as

\begin{equation}
\label{eqn:codeword_rep}
c(\acute{\boldsymbol{x}}) \triangleq \sign \left( \mathbb{E}[\, \acute{\boldsymbol{y}} - \boldsymbol{\tau} \mid \acute{\boldsymbol{x}}, \boldsymbol{H}, \boldsymbol{\tau} \,] \right) 
= \sign \left( \boldsymbol{H} \acute{\boldsymbol{x}} - \boldsymbol{\tau} \right),
\end{equation}

\noindent where $\acute{\boldsymbol{y}}$ is the unquantized observation vector obtained when $\acute{\boldsymbol{x}}$ is sent. The following is defined for ease of notation:

\begin{equation}
\label{eqn:hamm_dist}
\mathscr{H}(\acute{\boldsymbol{x}}) \triangleq \dham (\boldsymbol{r}, c(\acute{\boldsymbol{x}})),
\end{equation}

\noindent where $\dham \{\,.\,,\,.\,\}$ denotes the Hamming distance between its arguments. Now, the aim is to create a set $\mathcal{X} \subseteq \tilde{\mathcal{X}}$ whose cardinality is a hyperparameter denoted as $P$ and it consists the symbol-by-symbol detected vector $\check{\boldsymbol{x}}$ whose $k^{\textrm{th}}$ element is $\check{x}_k = \argmin_{x \in \mathcal{M}} |\tilde{x}_k - x|$ for $k=1, 2, \hdots, 2K$, and the $P-1$ candidate vectors from $\tilde{\mathcal{X}}$ such that the Hamming distance between their spatial codeword representations and $\boldsymbol{r}$ are the smallest. Including the symbol-by-symbol detection result as a candidate is enforced due to favorable performance, especially at low SNR. Next, in order to sort the candidate vectors according to the Hamming distance between their spatial codeword representations and the quantized observation vector, an ordered vector sequence is defined as

\begin{equation}
\label{eqn:candidate_sequence}
(\acute{\boldsymbol{x}}_i)_{i=1}^{|\mathcal{A}|} \in \tilde{\mathcal{X}} \setminus \{\check{\boldsymbol{x}}\} \: \textrm{such that} \: \mathscr{H}(\acute{\boldsymbol{x}}_{i}) \leq \mathscr{H}(\acute{\boldsymbol{x}}_{i+1}).
\end{equation}

Sequence indexing is a useful way to find the first $P-1$ candidates. Indexing may not be unique and encountering a situation such as $\mathscr{H}(\acute{\boldsymbol{x}}_{P-1}) = \mathscr{H}(\acute{\boldsymbol{x}}_{P})$ is possible. However, the performance gain obtained by the second stage increases as SNR increases, and for large enough $K$, a situation such as $\mathscr{H}(\boldsymbol{x})=\mathscr{H}(\acute{\boldsymbol{x}}_{P-1})$ is not likely. The final set $\mathcal{X}$ can be found as

\begin{equation}
\label{eqn:final_set}
\mathcal{X} = \{\acute{\boldsymbol{x}_1},\acute{\boldsymbol{x}}_2,\hdots,\acute{\boldsymbol{x}}_{P-1}\} \cup \{\check{\boldsymbol{x}} \}.
\end{equation}

Finally, ML detection (\ref{eqn:ml_detector_long}) is applied on the reduced set $\mathcal{X}$ instead of $\mathcal{M}^{2K}$ to obtain the second-stage solution.

\section{Pseudo-Random Quantization (PRQ) Scheme} \label{sec:prq_scheme}

\subsection{Proposed Quantization Strategy}

\begin{figure*}[tb]
\centering
\subfloat[Conventional Dithering]{\includegraphics[width=\columnwidth*7/20]{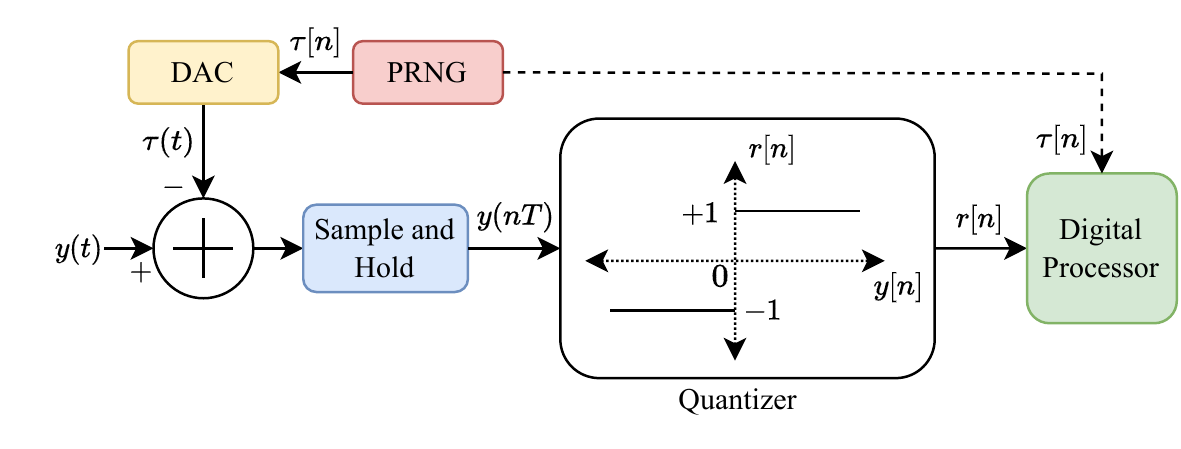}} \label{subfig:dither_v1}
\centering
\subfloat[Proposed Dithering]{\includegraphics[width=\columnwidth*7/20]{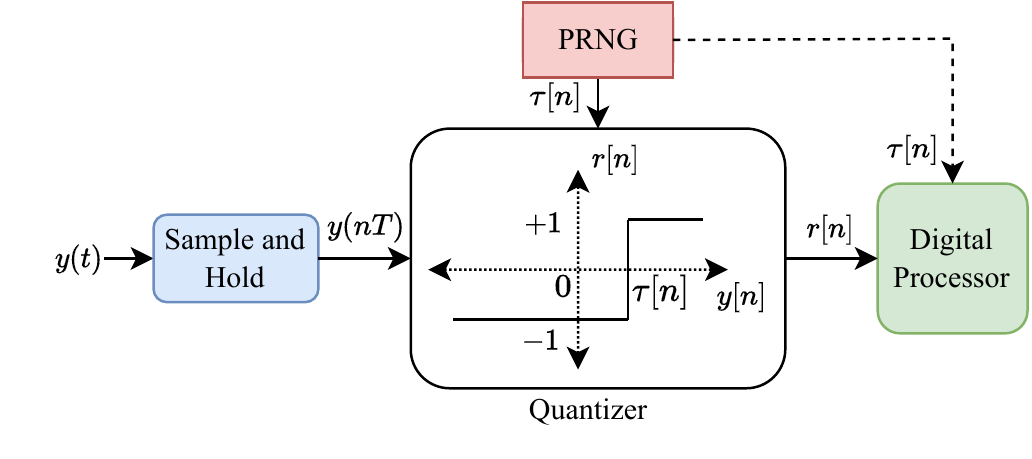}\label{subfig:dither_v2}}
\hfill
\caption{Illustrations of random (without the dashed line connection) and pseudo-random (with the dashed line connection) quantization schemes with the conventional (a) and the proposed (b) dithering architectures.}
\label{fig:dithering}
\end{figure*}

Generating artificial noise with a particular distribution is one way of dithering the input signal. However, the same dithering effect can be achieved by shifting the quantization thresholds. Dithering can also be called randomized quantization, with both random and pseudo-random \cite{dafsp} applications. In Fig. \ref{subfig:dither_v1}, the conventional version of dithering is illustrated where a randomly generated analog signal is subtracted from the incoming analog signal with the help of an additional DAC. PRNG is a pseudo-random number generator and helps obtain random sequences from a selected probability distribution. As seen in the figure, the advantage of PRQ is that we can utilize the knowledge of the dither signal/threshold during the detection operation. However, employing an additional DAC in the RF chain can be undesirable.

To obtain a more efficient system, we can rely on variable-threshold quantizers to help get the same effect as generating the dither signal and subtracting it from the received signal. The resulting setup is shown in Fig. \ref{subfig:dither_v2}. As seen in the figure, we rely on changing the quantization threshold instead of using an additional DAC. However, in practice, the response time of the ADC can become a burden if the threshold is varied at every symbol period or even every frame. Hence, keeping the thresholds constant during channel coherence time would be preferable. Since we deal with massive MIMO with large antenna arrays, the distribution of the dither signals, i.e., quantization thresholds is important rather than their different realizations due to the law of large numbers (LLN). Hence, we change the domain of dithering from temporal (time) to spatial (antennas) with this approach. From the circuit design perspective, since one-bit ADCs operate with basic comparator units that compare two voltage values to output a digital value, it should not be a challenging issue to compare the incoming voltage to a non-zero voltage value that can be set easily. Moreover, this is a more challenging issue for high-resolution ADCs where the quantization thresholds are adjusted continuously with the help of AGCs depending on the received signal power.

We propose a new quantization scheme for uplink one-bit massive MIMO systems to overcome the adverse effects of the SR phenomenon at high SNR, where either the performance peaks at a unique SNR value or a performance saturation occurs after a finite SNR. When a dithering scheme is applied in the system, the distribution of the dither signal, the threshold values for our scenario, must be selected. By observing different scenarios, we saw that selecting the Gaussian distribution is a useful technique. Then, the mean and variance of the thresholds must also be selected. Since we are trying to detect signals with zero-mean and our unquantized observations are zero-mean, the intuitive selection is to generate the threshold with zero-mean. The selection of the variance parameter is more challenging than the mean since it is hard to analytically track the effects of the empirical distribution of a pseudo-random selection. If the variance is too small, the performance is not affected. If it is too large, the performance gets worse since the threshold values become too large to differentiate relatively small amplitude differences. After many trials and observations, we saw that increasing $N$ requires the variance to be increased, whereas increasing $K$ requires the variance to be decreased. Also, the PRQ approach does not yield any benefits and causes performance degradation below a particular SNR value. Hence, gradually increasing the threshold variance after a threshold SNR value $\rho_{t}$ is an appropriate technique. In order to avoid selecting different variances for different scenarios, we parameterize the threshold SNR value as

\begin{equation}
    \label{eqn:threshold_snr}
    \rho_{t} = 5\log_{10} \left( 100\frac{K}{N} \right) \textrm{ dB}.
\end{equation}

\begin{figure*}[b]
\centering
\subfloat[Threshold Variance]{\includegraphics[width=\columnwidth*7/20]{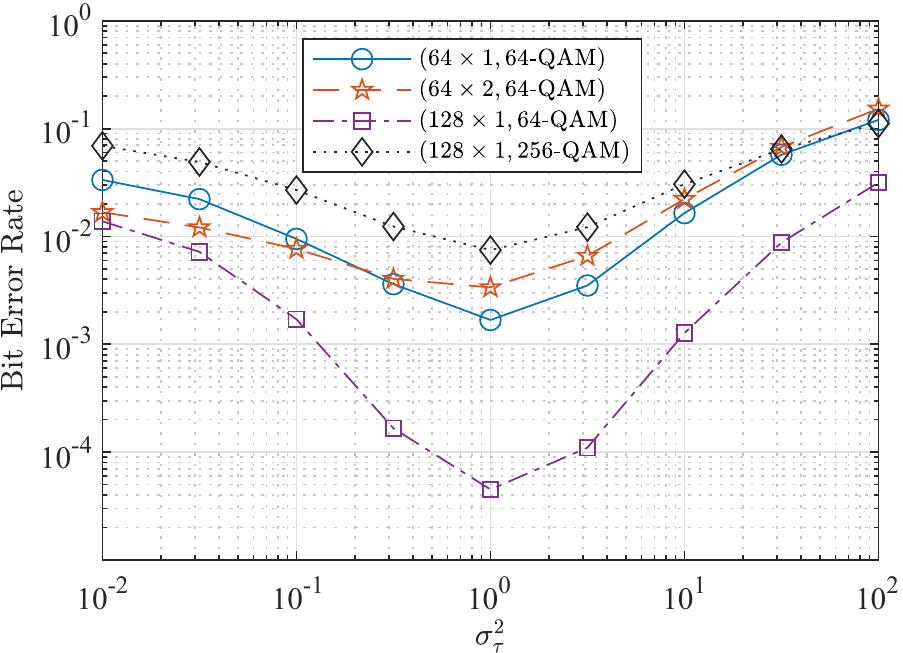} \label{subfig:threshold_variance_ml}}
\centering
\subfloat[SNR]{\includegraphics[width=\columnwidth*7/20]{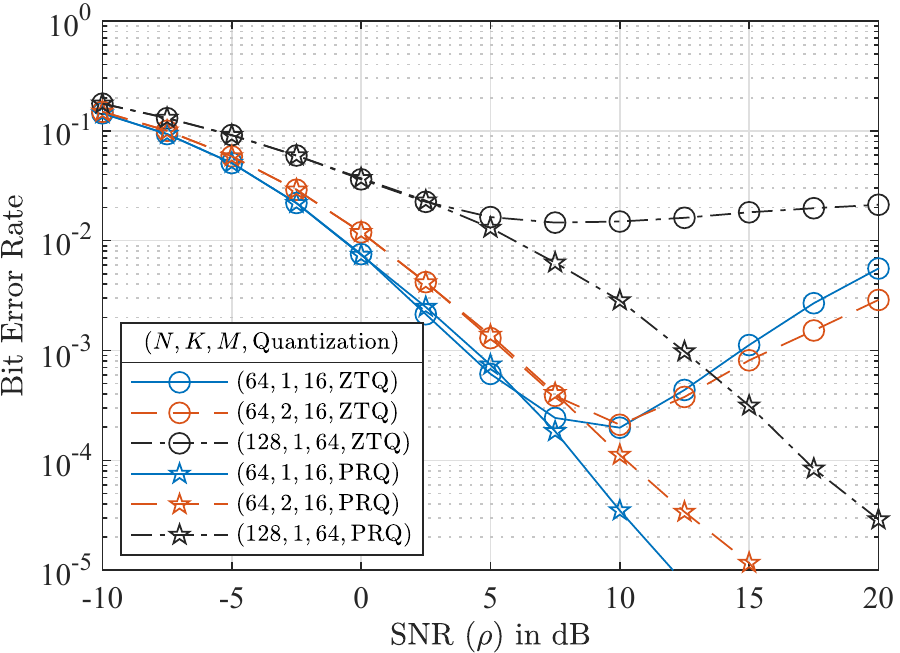} \label{subfig:snr_ml}}
\hfill
\caption{BER performance of the ML detector with respect to threshold variance at $\rho=20$ dB (a) and SNR with ZTQ and PRQ performances according to the chosen threshold SNR from (\ref{eqn:threshold_snr}).}
\label{fig:prq_effects}
\end{figure*}

This selection may not be optimal and relies only on empirical findings and observations. Now that the threshold SNR value is selected, the quantization threshold variance can be found accordingly using

\begin{equation}
    \label{eqn:threshold_variance}
    \sigma_{\tau}^2 = \R \left(\frac{E_s}{\rho_{t}} - N_0\right) = \R \left( \frac{1}{\rho_{t}} - N_0 \right).
\end{equation}

Upon generation, rescaling the thresholds to strictly satisfy $\lVert \boldsymbol{\tau} \rVert^2 = N \sigma_{\tau}^2$ is helpful to obtain more stable results. The proposed PRQ scheme does not require updates for different channel realizations and depends only on the SNR. In Fig. \ref{subfig:threshold_variance_ml} the effect of changing the threshold variance is investigated for different system setups using the ML detector in Rayleigh fading channel at a fixed SNR value. As can be seen, the threshold variance becomes an important factor for which a very high or a very low selection yields poor performance. Note that even though the selection $\sigma^2_{\tau}=1$ seems to be yielding the best performance, the difference between different scenarios become more evident when smaller SNR steps are used during simulations and when the number of users is selected to be higher. Note that from this point on, the threshold variance selections are made according to (\ref{eqn:threshold_variance}). In Fig. \ref{subfig:snr_ml}, the BER curves obtained with different system setups are shown both with ZTQ and PRQ in the Rayleigh fading channel. The peak performance is achieved, i.e., SR occurs, between 5-10 dB of SNR for each system with ZTQ. The two-user performance with $16$-QAM is better than that of the single-user at high SNR with ZTQ, which shows how multi-user interference (MUI) can act as a dither source. With PRQ, we not only stop the performance degradation but also obtain superior performance while approaching and at the SR point.

\subsection{Achievable Rate Analysis in One-Bit SIMO}

We have seen that the thresholds should be optimized according to the given channel realization for the conventional MIMO systems, i.e., when the number of antennas is small. Thankfully, massive MIMO systems can benefit from a more robust approach due to a large number of antennas, and important performance gains can be obtained without optimizing the thresholds and utilizing PRQ. Threshold optimization is also tricky since the mutual information or the error probability should be optimized. For our setup, the analytical expression for the conditional mutual information can be written as

\begin{multline}
    \label{eqn:mut_inf}
    \mathcal{I}(\boldsymbol{r}; \boldsymbol{x}\mid \boldsymbol{H}, \boldsymbol{\tau}) = \frac{-1}{M^K} \sum_{\boldsymbol{x}_1 \in \mathcal{M}^{2K}} \sum_{\boldsymbol{r} \in \{\pm 1\}^{2N}} \p(\boldsymbol{r} \mid \boldsymbol{x}_1, \boldsymbol{H}, \boldsymbol{\tau}) \log_2 \left( \frac{1}{M^K} \sum_{\boldsymbol{x}_2 \in \mathcal{M}^{2K}} \p(\boldsymbol{r} \mid \boldsymbol{x}_2, \boldsymbol{H}, \boldsymbol{\tau}) \right) \\ - \frac{1}{M^K} \sum_{\boldsymbol{x}_3 \in \mathcal{M}^{2K}} \sum_{n=1}^{2N} \mathcal{H}_b \left( \Phi \left( \sqrt{2\rho} ( \boldsymbol{h}_n^T \boldsymbol{x}_3 - \tau_n ) \right) \right),
\end{multline}

\noindent where $\mathcal{H}_b(.)$ is the binary entropy function. A detailed derivation of mutual information is given in Appendix \ref{app:mut_inf}. Finding the optimal thresholds is difficult since the solution is not necessarily unique, and the dimensionality is very high for the massive MIMO setup. To observe the effects of PRQ, we resort to random sampling by utilizing the Monte Carlo method. For the simulations, we aim to find the achievable rate with ZTQ and PRQ in a SIMO system where a BS equipped with $4$ antennas is serving a single user in AWGN and Rayleigh fading channels. The conditional mutual information for any given channel realization is calculated for $500$ random realizations of the thresholds for the PRQ scheme and $1$ with zero thresholds for ZTQ. The average rate and the maximum rate obtained during trials for each channel are recorded with PRQ. Since the AWGN channel is deterministic, only a single channel realization is utilized. For the Rayleigh channel, the average is calculated over $1000$ channel realizations.

\begin{figure}[b]
\centering
\subfloat[AWGN Channel]{\includegraphics[width=\columnwidth*7/20]{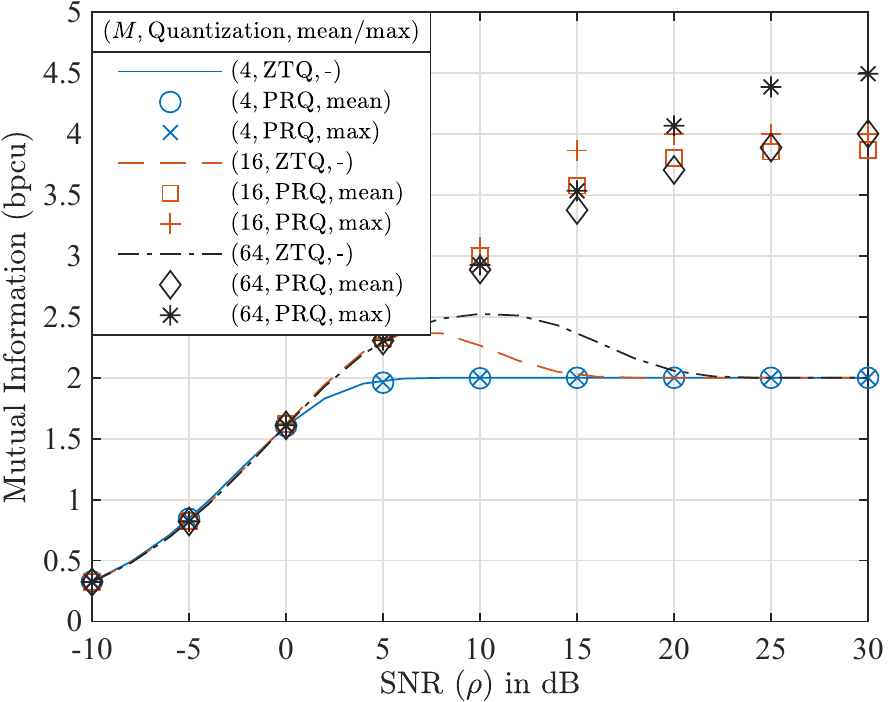} \label{subfig:mi_awgn}}
\centering
\subfloat[Rayleigh Channel]{\includegraphics[width=\columnwidth*7/20]{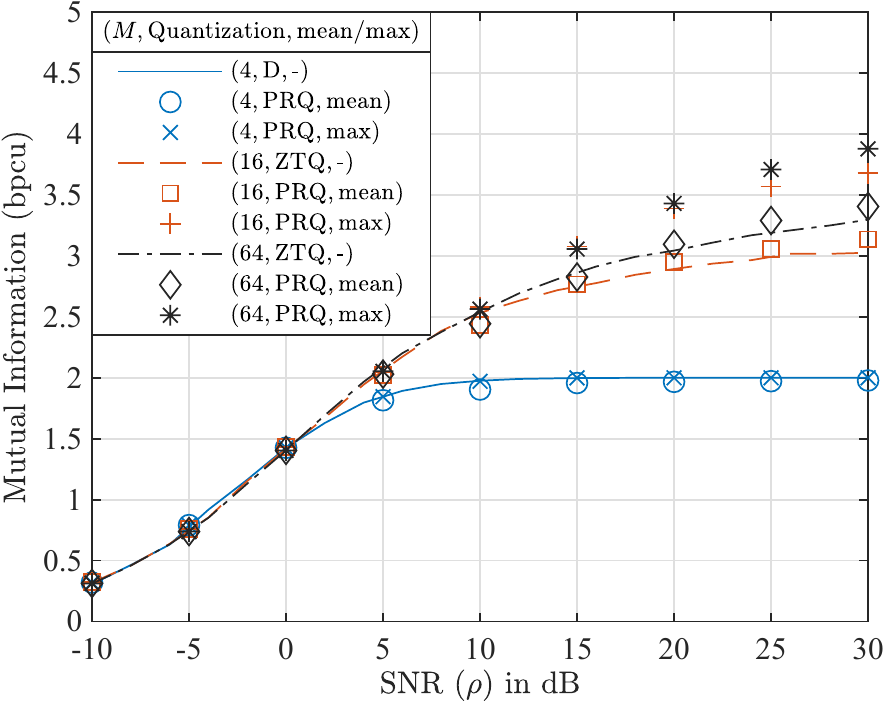} \label{subfig:mi_rayl}}
\caption{Mutual information plotted against SNR for both AWGN (a) and Rayleigh (b) channels where $N=4$, $K=1$, and $M=4,16,64$ with one-bit ZTQ (lines) and PRQ (markers) schemes. The average and the maximum rate obtained with the PRQ scheme are recorded.}
\label{fig:mut_inf}
\end{figure}

The mutual information plots with respect to SNR are shown in Fig. \ref{subfig:mi_awgn} for the AWGN channel and in Fig. \ref{subfig:mi_rayl} for the Rayleigh fading channel. The performances in both channels exhibit no significant difference when the QPSK constellation is used. This is an intuitive result since each pair of ADCs from each antenna divides the 2D space into four regions, the cardinality of the QPSK alphabet is also four. However, the behavior changes drastically for 16-QAM and 64-QAM. The SR phenomenon is visible in the AWGN channel with $16$-QAM and $64$-QAM constellations. We can no longer observe SR in this $4$-antenna scenario for the Rayleigh fading channel. This seems to be due to the amplitude and phase-varying nature of the Rayleigh fading channel, which helps obtain different instantaneous SNR values at each branch. Nonetheless, the PRQ scheme outperforms the ZTQ scheme for both channel types, even on average. If the thresholds are optimized with respect to the given channel realization, the rate can be increased further in this low-dimensional setup.

\subsection{Minimum Hamming Distance Analysis}

Calculating mutual information for the massive MIMO setup is not feasible due to very large dimensionality. Therefore, another approach is followed to grasp how the PRQ scheme works for massive MIMO systems using an intuition from coding theory \cite{Hamm_Dist_Decod}. Quantized observations can be seen as binary codewords in space for each given channel realization. We can only intervene in the encoding process by changing the thresholds. The resultant spatial channel code will determine our error performance. A measure of the performance of a code is its minimum Hamming distance, which determines the diversity order of a code in fading channels \cite{ss_diversity}. In this part, we will investigate the average of the minimum Hamming distance of the space code obtained in the Rayleigh fading channel via Monte Carlo simulations at infinite SNR with both ZTQ and PRQ schemes. The infinite SNR scenario is chosen to understand the high SNR behavior, which is a critical part of our purposes. For a given channel and a given threshold realization, the minimum Hamming distance of the codebook is calculated for the infinite-SNR scenario as

\begin{equation}
\label{eqn:min_ham_dist_codebook}
d_\mathrm{H}^{\,\mathrm{min}} = \min_{\boldsymbol{x}_1, \boldsymbol{x}_2 \in \mathcal{M}^{2K}} \dham(c(\boldsymbol{x}_1), c(\boldsymbol{x}_2)),
\end{equation}

\noindent where $c(.)$ is the spatial codeword representation of its argument vector, as in (\ref{eqn:codeword_rep}).

\begin{table}[tb!]
\centering
\caption{Average Minimum Hamming Distance of the Space Code in the Rayleigh Fading Channel}
\label{tab:min_hamm_dist}
\resizebox{\columnwidth/4}{!}{%
\begin{tabular}{|c|c|c|c|c|}
\hline
$N$ & $K$ & $M$  & with ZTQ & with PRQ \\ \hline
64  & 1 & 4  & 64.00   & 47.59    \\ \hline
64  & 2 & 4  & 35.41   & 28.98    \\ \hline
64  & 1 & 16 & 0.00    & 11.92    \\ \hline
64  & 2 & 16 & 0.00    & 5.03     \\ \hline
64  & 1 & 64 & 0.00    & 1.85     \\ \hline
128 & 1 & 16 & 0.00    & 25.63    \\ \hline
128 & 1 & 64 & 0.00    & 6.39     \\ \hline
\end{tabular}%
}
\end{table}

The average of the minimum Hamming distance in the Rayleigh fading channel is obtained for both ZTQ and PRQ scenarios where $10^4$ randomly generated channel realizations are matched with randomly generated quantization thresholds. Table \ref{tab:min_hamm_dist} shows the results for different scenarios. An interesting result obtained from these simulations is that when ZTQ is employed with $M \geq 16$, the minimum Hamming distance of the space code is $0$, which means that the code is not uniquely decodable. Errors will indeed occur during the decoding operation due to the SR phenomenon; however, if we utilize the PRQ scheme, which gives us a chance to intervene in the codebook design, the minimum Hamming distance of the code increases. Hence, a superior performance compared to the ZTQ scheme will be obtained. QPSK modulation does not benefit from PRQ, as was the case during the mutual information calculations. Also, a larger alphabet size leads to a smaller minimum Hamming distance, whereas a larger number of antennas, i.e., larger codeword length, leads to an increased minimum Hamming distance for the space code with PRQ. Since the rate of the code is $R_c=\frac{K\log_2(M)}{2N}$, increasing $K$ or $M$ results in a lower, and increasing $N$ leads to a larger minimum Hamming distance.

\section{Computational Complexity Analysis}

Even though the gradient descent does not require matrix inversion, Newton's method requires much fewer iterations to reach the optimum \cite{Newtons_Method, optimization_comparison}. The part that dominates the complexity of BND is the calculation of the step towards the optimum. Calculating the Hessian matrix dominates the complexity with $\mathcal{O}(NK^2)$ since $N \gg K$ for massive MIMO systems. Since this procedure is repeated until the termination of the algorithm, the resulting complexity is $\mathcal{O}(NK^2T)$, where $T$ denotes the number of iterations. Note that computation of the nonlinear function $\psi(.)$ can be made using only $\varphi(.)$ since they share the same arguments. The average number of iterations of BND is around $3-5$, and the number of unreliable indices for NCD is generally small at high SNR. Therefore, the complexity-dominant part of NCD is the last part where ML detection is applied on the reduced set with cardinality $P$. Even though we let $P$ get as large as $2K$ during simulations, the average cardinality of the candidate set is small at high SNR, resulting in a complexity of $\mathcal{O}(NK^2)$ also for NCD.

\begin{table}[tb]
\centering
\caption{Computational Complexity Analysis}
\label{tab:comp_complexity}
\resizebox{\columnwidth/2}{!}{%
\begin{tabular}{|c|c|c|c|}
\hline
\textbf{Method}     & \textbf{Preprocessing}              & \textbf{Stage 1}                   & \textbf{Stage 2}           \\ \hline
\textbf{BND-NCD} & -                                   & $\mathcal{O}(NK^2T)$               & $\mathcal{O}(NK^2)$        \\ \hline
\textbf{MRC/BMRC}  & $\mathcal{O}(NK)$   & $\mathcal{O}(NK)$          & -                    \\ \hline
\textbf{ZF/BZF}    & $\mathcal{O}(NK^2)$ & $\mathcal{O}(NK)$          & -                    \\ \hline
\textbf{ML}        & $\mathcal{O}(NK|\mathcal{M}|^{2K})$ & $\mathcal{O}(N|\mathcal{M}|^{2K})$ & -                          \\ \hline
\textbf{SVM-based} & -                   & $\mathcal{O}(NK\kappa(N))$ & -                    \\ \hline
\textbf{OBMNet-NNS} & Offline Training                    & $\mathcal{O}(NKL)$                 & $\mathcal{O}(\max(M,N)KM)$ \\ \hline
\end{tabular}%
}
\end{table}

The complexities of the proposed BND-NCD and the existing detectors from the literature are shown in Table \ref{tab:comp_complexity}. MRC-based detectors offer the lowest complexity with poor error performance. Then comes ZF-based detectors that offer better but not adequate performance and with higher computational complexity due to matrix inversion. The ML detector has the highest complexity, which is not feasible to operate in commercial systems. SVM-based detector from \cite{SVM} reportedly has complexity $\mathcal{O}(NK\kappa(N))$, where $\kappa(.)$ is a super-linear function, and large values of $N$ may lead to unaffordable complexity. OBMNet from \cite{OBMNet}, reportedly has complexity $\mathcal{O}(NKL)$ in its first stage, where $L$ is the number of layers of the proposed deep neural network (DNN) architecture, and complexity of $\mathcal{O}(\max(M, N)KM)$ in its second stage nearest neighbor search (NNS) algorithm. Even though OBMNet does not require matrix inversion, the neural network needs an offline training stage, and a much fewer number of iterations is required by BND, which makes the complexities of the detectors comparable. Also, the algorithm can be terminated early when BND's stopping criterion is satisfied. In contrast, the number of layers defined as part of the deep unfolded network architecture is a constant for OBMNet. Finally, the second stage NNS algorithm of \cite{OBMNet} is a recursive algorithm with many for-loops. NCD omits for-loops and calculates the candidate set efficiently.

\section{Simulation Results} \label{sec:num_results}

In this section, the error performance of the proposed methods is investigated. For BND, the maximum number of iterations $T_{\mathrm{max}}=15$, the damping factor $\zeta$ is calculated by taking $\rho_d=20-160\frac{K}{N}$ dB, the penalty control parameter $\theta=20$, and the termination threshold $\epsilon=\frac{10^{-4}}{\log_2(M)}$. For NCD, the maximum size of the set of nearest spatial codewords $P=2K$, to adjust the size of the unreliable region, $\gamma=d/2$ with $d=\sqrt{\frac{3}{2(M-1)}}$, and the maximum size of set of unreliable indices $U_{\mathrm{max}}=K$.

\subsection{Verification of the Convergence of BND}

\begin{figure*}[t]
\centering
\subfloat[BER]{\includegraphics[width=\columnwidth*7/20]{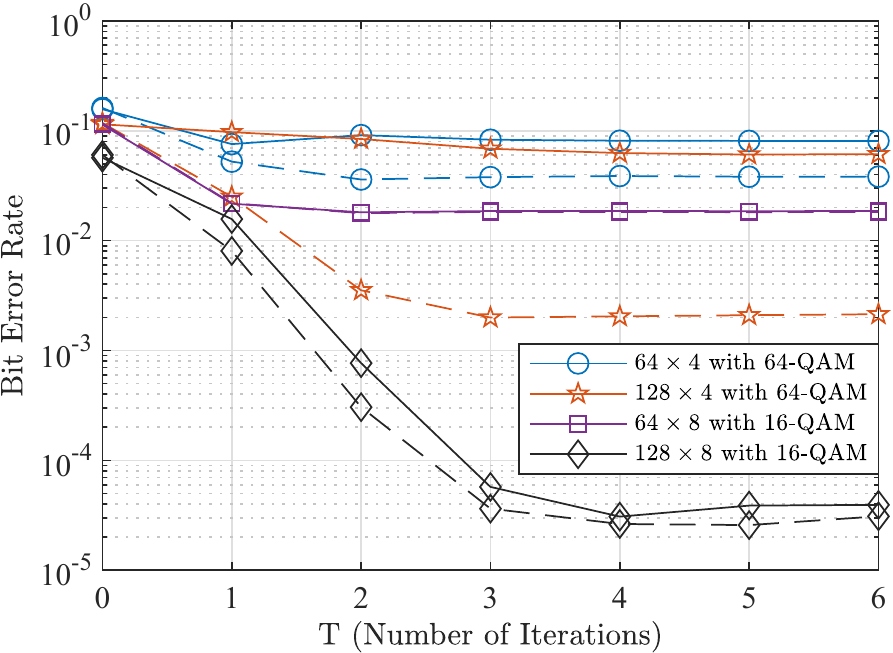} \label{subfig:convergence_ber}}
\centering
\subfloat[Cost Function (\ref{eqn:unconstrained_opt})]{\includegraphics[width=\columnwidth*7/20]{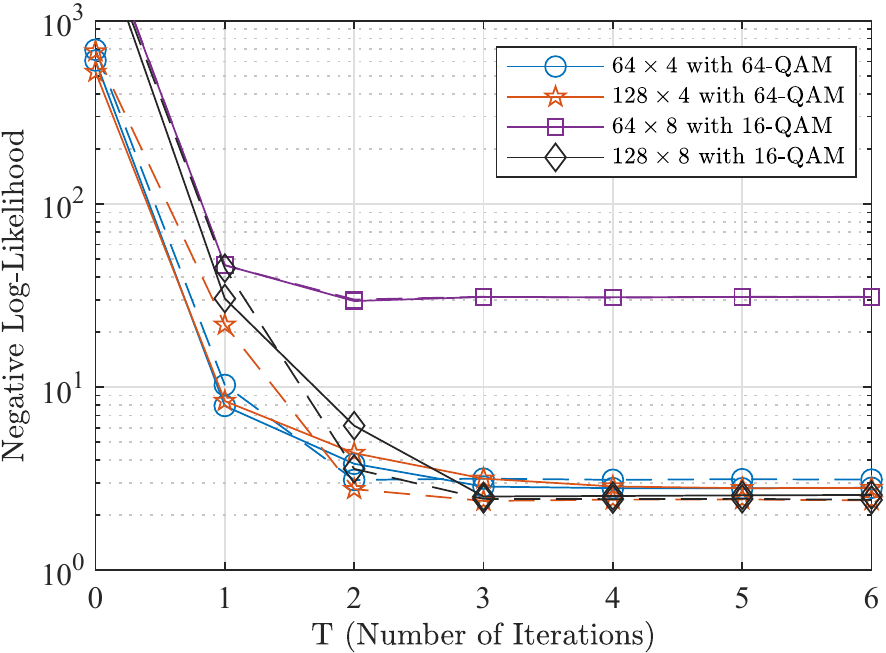} \label{subfig:convergence_costf}}
\hfill
\caption{Empirical convergence analysis using the BER performance (a) and negative of the cost function (b) to be optimized with respect to number of iterations of the proposed first stage detector BND at $\rho=30$ dB with different system setups where solid lines reflect the results obtained with ZTQ and dashed lines are for PRQ.}
\label{fig:convergence}
\end{figure*}

We start by verifying the convergence of the proposed first stage, BND. In Fig. \ref{fig:convergence}, the error performance (a) and the negative of the cost function (b) from (\ref{eqn:unconstrained_opt}) with respect to the number of iterations are plotted for different system setups with both ZTQ and PRQ. As can be seen in the plots, the convergence time depends on the system setup. However, the algorithm converges within 2-4 iterations for each scenario.

\subsection{One and Two-Stage ZTQ and PRQ Performances}

\begin{figure*}[b]
\centering
\subfloat[ZTQ]{\includegraphics[width=\columnwidth*7/20]{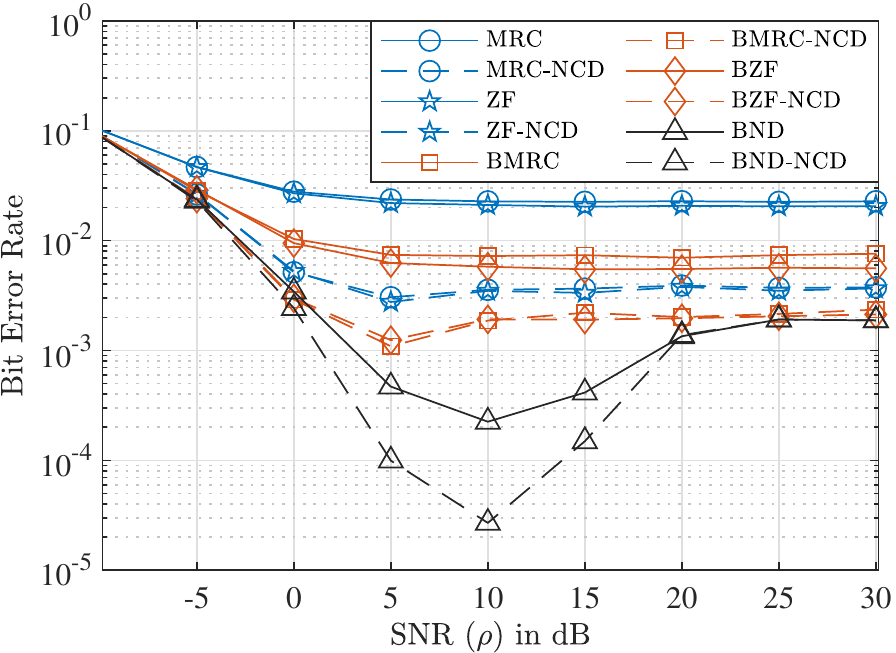} \label{subfig:linear_comparison_ztq}}
\centering
\subfloat[PRQ]{\includegraphics[width=\columnwidth*7/20]{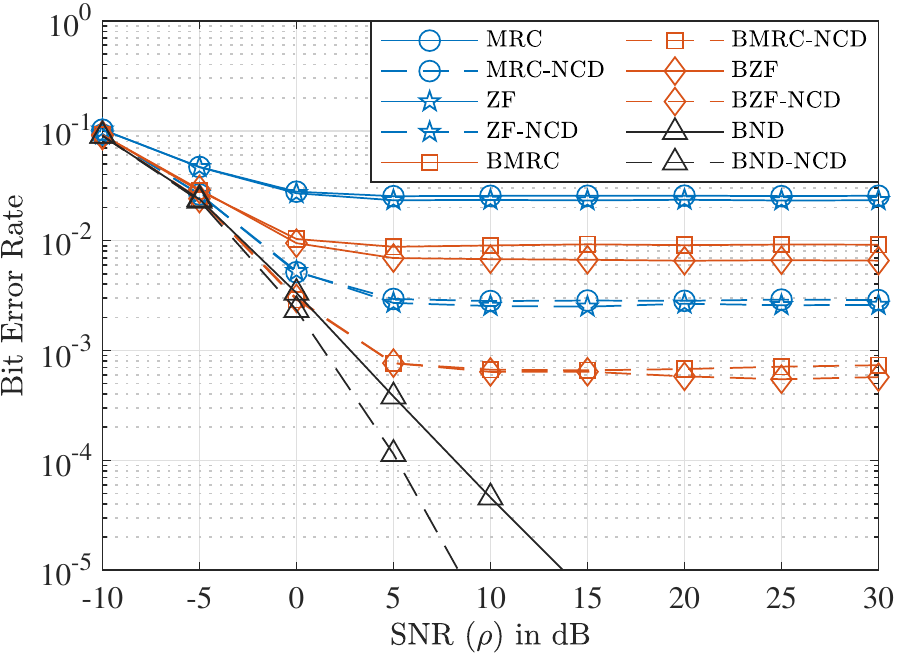} \label{subfig:linear_comparison_prq}}
\hfill
\caption{One and two-stage BER performances of the linear and the proposed detectors in a $128 \times 4$ system with $16$-QAM both with ZTQ (a) and PRQ (b) schemes, where linear detectors are also matched with the proposed NCD for the second stage.}
\label{fig:linear_comparison}
\end{figure*}

We now examine he one-stage and two-stage performances of the linear detectors and the proposed BND-NCD. The linear detectors are also matched with the proposed NCD at the second stage. The resultant BER curves obtained by ZTQ and the proposed PRQ schemes are shown in Fig. \ref{fig:linear_comparison}. BND outperforms the linear detectors, ZF-type receivers top their MRC-type counterparts, and Bussgang-based receivers perform slightly better than the conventional linear receivers. The performance gap between the conventional and Bussgang-based linear detectors is smaller than the previously reported works from the literature \cite{ADMM, OBMNet} due to the quantization label $\ell_q$. Substantial gains can be obtained by utilizing a two-stage approach for all scenarios. The linear detectors do not benefit much from PRQ, and only the second stage performance of BZF differs where a slightly lower error floor is obtained. SR is visible for all detectors with ZTQ, whereas it is no longer observed for BND and BND-NCD with PRQ.

\subsection{Effect of Changing the Number of Users and the Number of BS antennas}

\begin{figure*}[t]
\centering
\subfloat[Number of Users]{\includegraphics[width=\columnwidth*7/20]{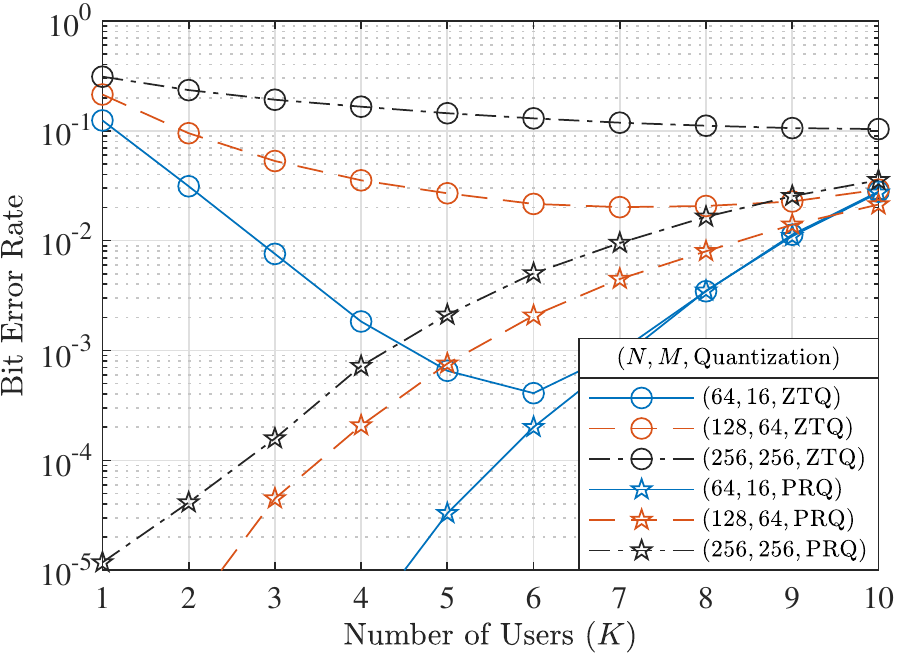} \label{subfig:number_of_users}}
\centering
\subfloat[Number of BS Antennas]{\includegraphics[width=\columnwidth*7/20]{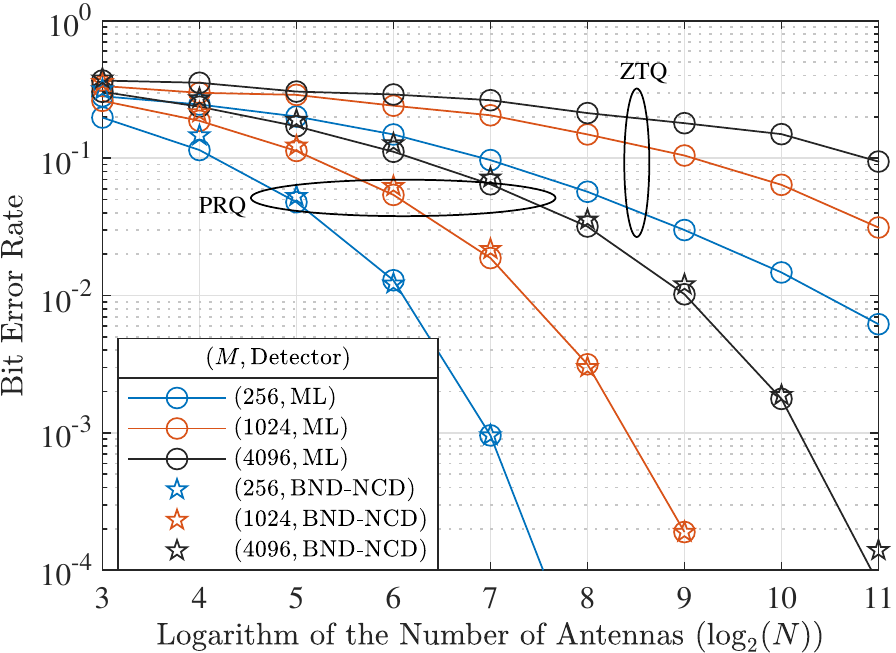} \label{subfig:number_of_antennas}}
\hfill
\caption{BER performance of BND-NCD obtained by ZTQ and PRQ with respect to the number of users (a) and the number of BS antennas for a SIMO system (b) where $\rho=30$ dB with different system setups. For the second plot (b) only, $U_{\mathrm{max}}=2$ and $P=\log_2(M)/2$.}
\label{fig:changing_parameters}
\end{figure*}

Next, we investigate the effect of the number of users on the system's performance at a fixed high-SNR scenario with the proposed BND-NCD by utilizing both ZTQ and PRQ in Fig. \ref{subfig:number_of_users}. Starting with $K=1$, increasing the number of users results in better performance up to a certain point with ZTQ, which also shows how the multi-user scenario helps obtain dithering. As the number of users increases, the performance gap between the ZTQ and PRQ schemes decreases and they perform the same after some point. As $N$ increases, the starting point where ZTQ and PRQ schemes perform the same is delayed to a larger value of $K$, which means that the merits of the PRQ scheme get more evident as the number of BS antennas increases.

The performance of SIMO systems that employ ML and BND-NCD with various high-order modulations are obtained with the ZTQ and PRQ schemes against the logarithm of the number of BS antennas in Fig. \ref{subfig:number_of_antennas}. Based on this and the previous results, we can state that when $\sfrac{K}{N}$ is small, the spatial diversity of the system is not adequately utilized with the ZTQ scheme. Moreover, we can also see that modulation orders such as $256$, $1024$, and $4096$ can be used in one-bit quantized systems with PRQ if a sufficient number of BS antennas are utilized. The proposed two-stage detector can perform very close to ML with much lower computational complexity.

\subsection{Performance Comparison of the Proposed Method with the Existing Methods}

\begin{figure}[t]
\centering
\includegraphics[width=\columnwidth*7/20]{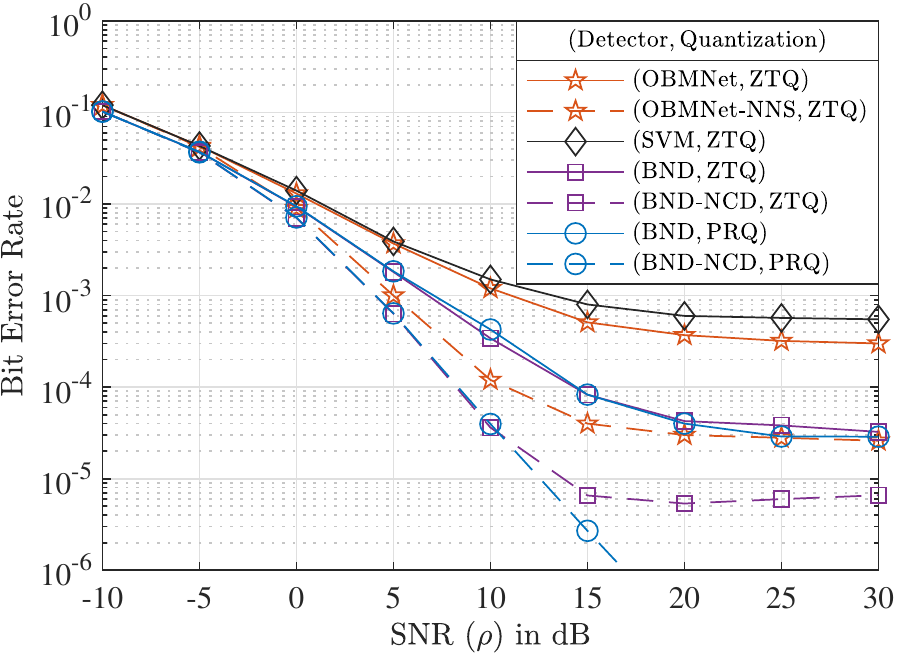}
\caption{BER performance comparison of the proposed detector with OBMNet \cite{OBMNet} and SVM-based \cite{SVM} detectors from the literature in a $128\times8$ system that employs $16$-QAM. NNS is the second-stage detector from \cite{OBMNet}. The maximum cardinality of the set of nearest neighbors for NNS is chosen as $2K=16$, the same as NCD.}
\label{fig:detector_comparison}
\end{figure}

The performances of the proposed detector and the existing detectors from the literature, namely, SVM-based from \cite{SVM}, and OBMNet from \cite{OBMNet} are compared in Fig. \ref{fig:detector_comparison}. The second stage for OBMNet is the nearest neighbor search (NNS) algorithm from \cite{OBMNet}, and the SVM-based detector is utilized as a single-stage detector. The cardinality of the set of nearest neighbors to conduct ML detection for the second stage detector of OBMNet is also chosen as $2K=16$. The plots show that the high-SNR error floor of the one-stage BND coincides with that of the two-stage OBMNet-NNS, the most recently proposed high-order modulation supporting detection scheme from the literature. Even when ZTQ is employed, BND-NCD outperforms both machine learning based approaches. Also, unlike any of the detectors included for comparison, the proposed two-stage BND-NCD with PRQ reduces the error floor below $10^{-6}$. BND requires $4.5$ iterations on average, whereas the number of layers of OBMNet, which is a DNN-based detector, is $15$.

\subsection{Error Performance and Complexity Analysis of the Proposed Methods in Large-Scale Systems}

\begin{figure*}[t]
\centering
\subfloat[BER Performance]{\includegraphics[width=\columnwidth*7/20]{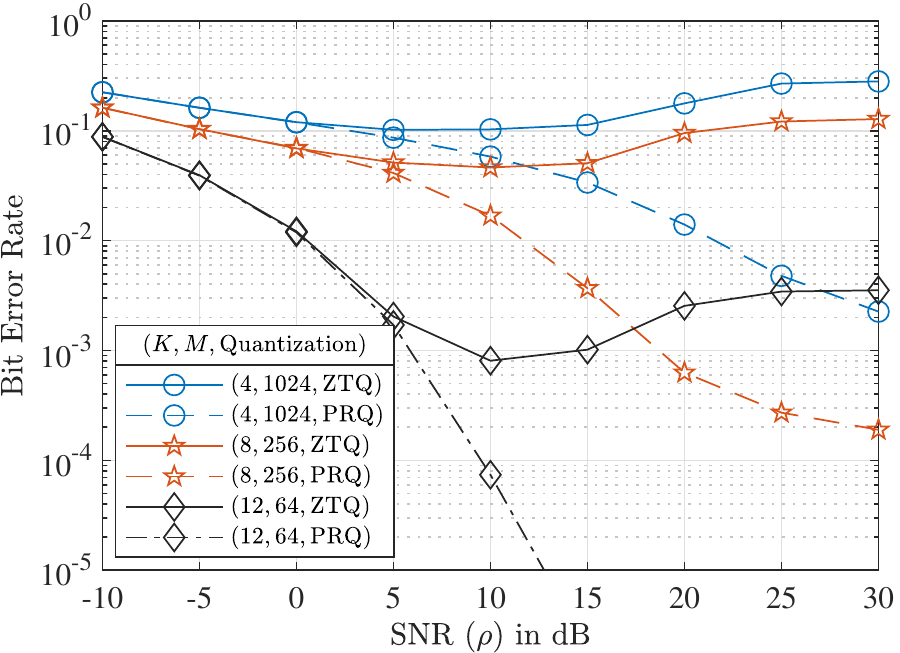} \label{subfig:ber_analysis}}
\subfloat[Computational Complexity]{\includegraphics[width=\columnwidth*7/20]{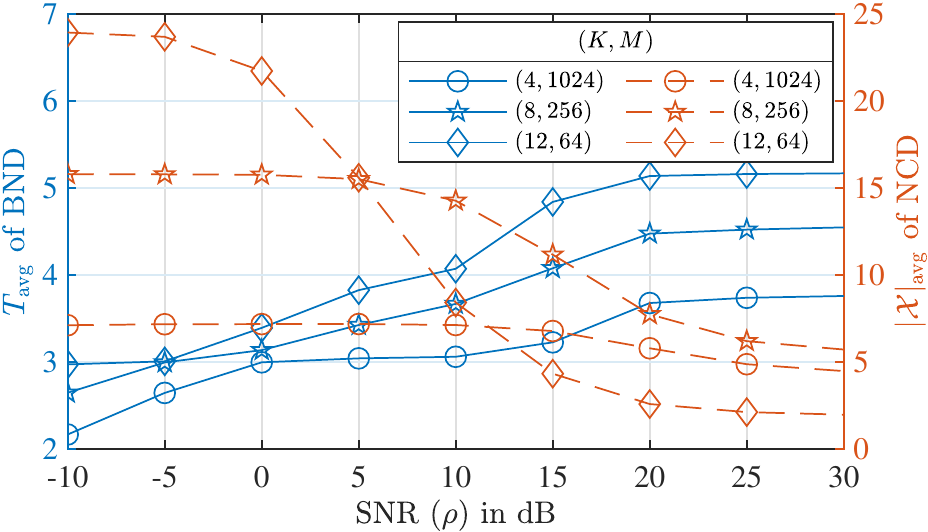} \label{subfig:cc_analysis}}
\hfill
\caption{BER performance (a) and complexity-related measurements (b) against SNR obtained with BND-NCD when $N=512$ for $K=4$ with $1024$-QAM, $K=8$ with $256$-QAM, $K=12$ with $64$-QAM. BER plots are obtained with ZTQ and PRQ, whereas complexity-related measurements are recorded with PRQ. $T_{\mathrm{avg}}$ is the average number of iterations of BND, and $|\mathcal{X}|_{\mathrm{avg}}$ is the average size of the reduced set in NCD.}
\label{fig:multiple_plots}
\end{figure*}

The error performances of several high-order modulation schemes in multi-user settings are examined. Additionally, complexity-related measurements regarding the average number of iterations of BND and the average size of the reduced set $\mathcal{X}$ are also made. BER performances (a) of different system setups with high-order modulations using ZTQ and PRQ, and complexity-related measurements (b) are plotted in Fig. \ref{subfig:cc_analysis} using PRQ. The proposed PRQ scheme can mitigate the effects of SR and lower the error floors. Hence, the sources of diversity are better exploited by PRQ. For complexity, we can see that $3-5$ iterations are required for BND, and the number of iterations increases as $K$ increases. $|\mathcal{X}|_{\mathrm{avg}}$ decreases as SNR increases, which means that employing only the first-stage BND is sufficient at low SNR since the performance gaps between the one and two-stage approaches are very small at low SNR, which was shown in Fig. \ref{fig:linear_comparison}. At high SNR, even though the number of users is the largest for $64$-QAM, it also has the smallest $|\mathcal{X}|_{\mathrm{avg}}$. Therefore, as $M$ increases, a more accurate estimation of the amplitudes is required. Note that the termination threshold $\epsilon$ of BND and the maximum size of the set of candidates $P$ of NCD can be changed to obtain a trade-off between error performance and computational complexity.

\subsection{Effect Imperfect CSI on the Performance of PRQ}

\begin{figure*}[tb]
\centering
\subfloat[ZTQ]{\includegraphics[width=\columnwidth*7/20]{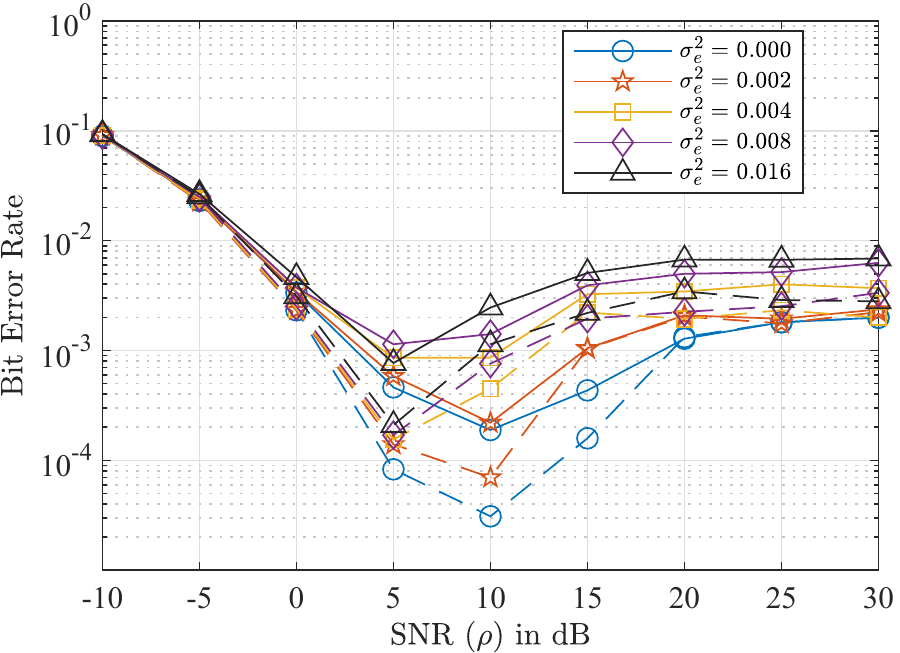} \label{subfig:imperfect_csi_ztq}}
\subfloat[PRQ]{\includegraphics[width=\columnwidth*7/20]{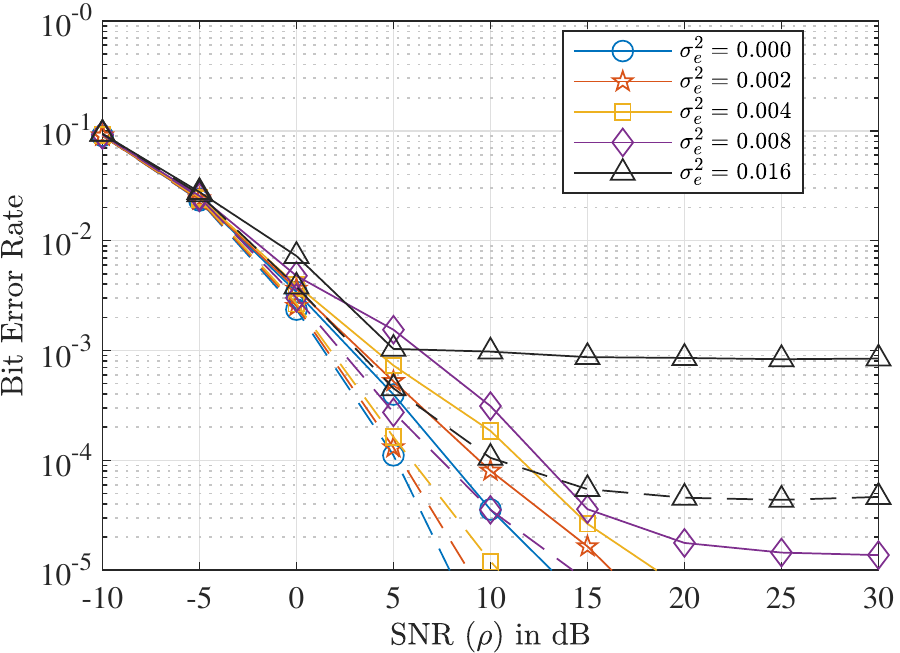} \label{subfig:imperfect_csi_prq}}
\hfill
\caption{BER performance of the proposed detector in one and two stages for the imperfect CSI scenario in a $128 \times 4$ system with $16$-QAM with ZTQ (a) and PRQ (b), where solid lines are for the one-stage BND and the dashed lines represent the two-stage BND-NCD approach.}
\label{fig:imperfect_csi}
\end{figure*}

In the final part, we investigate the effect of imperfect CSI knowledge at the BS on the detection performance. We model the channel estimate as the sum of the original channel $\bar{\boldsymbol{H}}$ and the estimation error $\Bar{\boldsymbol{E}}$ such that $\tilde{\bar{\boldsymbol{H}}} = \bar{\boldsymbol{H}} + \bar{\boldsymbol{E}}$. We assume that each element of the estimation error matrix $[\bar{\boldsymbol{E}}]_{(n,k)} \sim \mathcal{CN}(0,\sigma^2_e)$, to simply model the distortion. This model can be valid when PRQ is also used for channel estimation due to better amplitude recovery where the estimate can be obtained with a Gaussian distributed error \cite{Throughput_Analysis}, which is left as future work. In Fig. \ref{fig:imperfect_csi}, the BER performances obtained with ZTQ and PRQ for various estimation error variance scenarios are plotted using BND and BND-NCD with both ZTQ (a) and PRQ (b). As can be seen when the estimation error is not large, the performance is not affected much with PRQ. But even when the estimation error is large, PRQ still outperforms ZTQ, which shows the robustness of the PRQ and BND-NCD methods for the imperfect CSI scenario.

\section{Conclusion} \label{sec:conc}
This paper proposes a new PRQ scheme, modified linear detectors with PRQ, and a new two-stage detector for uplink one-bit massive MIMO systems. In the first stage, BND is utilized to estimate the transmitted signal using the log-likelihood function via Newton's Method with penalty. The second-stage detector, NCD, is used to refine the estimate from the first stage to enhance detection performance, which creates a small set of candidates based on the sign constraints imposed by one-bit quantization. The proposed PRQ scheme relies on changing the quantization thresholds to help mitigate the detrimental effects of SR on the performance. With the proposed two-stage detector and PRQ, one-bit multi-user massive MIMO systems can operate with high-order modulation schemes such as $256$-QAM or $1024$-QAM, and superior error performance to the existing detectors can be obtained with comparable complexity. We extend our work to frequency-selective channels in \cite{qnd} to analyze the effect of inter-symbol interference (ISI) with PRQ, and our detailed work on detection in massive MIMO with PRQ can be found in \cite{yilmaz_2023}.

\section*{Acknowledgments}
This study is partially supported by Vodafone Turkey within the framework of 5G and Beyond Joint Graduate Support Program coordinated by Information and Communication Technologies Authority of Turkey.

\appendices

\section{Computations of the Nonlinear Functions} \label{app:nonlinear_func}

Computation of $\ln(\Phi(x))$ can be a problem when $x$ is small, and divergence towards $-\infty$ for small arguments in finite precision can be encountered. We can approximate $\ln(\Phi(x))$ using its first-order derivative to avoid possible divergent behavior as

\begin{equation}
\label{eqn:logcdf_approx}
\ln(\Phi(x)) \cong \ln(\Phi(c)) + \varphi(c) (x-c),
\end{equation}

\noindent for $x<c \in \mathbb{R}$, where $\varphi(x)=\frac{d}{dx}\ln(\Phi(x))=\frac{\phi(x)}{\Phi(x)}$ is the first order derivative of $\ln(\Phi(x))$. $c=-38.2$ is a suitable choice for MATLAB's arithmetic calculations since divergent behavior is observed for smaller values. However, the first and second-order derivatives $\varphi(x)$ and $\psi(x)$ also show divergent behavior around the same point $c$. The asymptotic behavior of $\varphi(x)$ towards $-\infty$ can be found as

\begin{equation}
\label{eqn:fod_asymptotic}
\lim_{x \to -\infty} \frac{\phi(x)}{\Phi(x)} = \lim_{x \to -\infty} - \frac{x \phi(x)}{\phi(x)} = \lim_{x \to -\infty} -x,
\end{equation}

\noindent which means that for $x<c$, $\varphi(x)=-x$ and $\frac{d^2}{dx^2}\ln(\Phi(x))=\psi(x)=-1$ can be utilized in practice. Note that a look-up table can be generated to calculate nonlinear functions to decrease the complexity.

\section{Derivation of the Conditional Mutual Information}
\label{app:mut_inf}
Starting with the definition of conditional mutual information:

\begin{equation}
\label{eqn:mi_def}
\mathcal{I}(\boldsymbol{r}; \boldsymbol{x} \mid \boldsymbol{H}, \boldsymbol{\tau}) =
\mathcal{H}(\boldsymbol{r} \mid \boldsymbol{H}, \boldsymbol{\tau}) -
\mathcal{H}(\boldsymbol{r} \mid \boldsymbol{x}, \boldsymbol{H}, \boldsymbol{\tau}),
\end{equation}

\noindent where $\mathcal{H}(.)$ is the entropy function. The first term in (\ref{eqn:mi_def}) can be calculated by the definition of conditional entropy as

\begin{equation}
\label{eqn:mi_first}
\mathcal{H}(\boldsymbol{r} \mid \boldsymbol{H}, \boldsymbol{\tau}) =
- \sum_{\boldsymbol{r} \in \{\pm1\}^{2N}} \p(\boldsymbol{r} \mid \boldsymbol{H}, \boldsymbol{\tau}) \log_2(\p(\boldsymbol{r} \mid \boldsymbol{H}, \boldsymbol{\tau})),
\end{equation}

\noindent where $p(\boldsymbol{r} \mid \boldsymbol{H}, \boldsymbol{\tau})$ can be found by averaging the likelihood function $p(\boldsymbol{r} \mid \boldsymbol{x}, \boldsymbol{H}, \boldsymbol{\tau}) = \prod_{n=1}^{2N} \Phi \left( \frac{r_n (\boldsymbol{h}_n^T \boldsymbol{x} - \tau_n)}{\sqrt{N_0/2}} \right)$  over all possible $\boldsymbol{x}$ vectors such that

\begin{equation}
\label{eqn:expected_over_x_likelihood}
\p(\boldsymbol{r} \mid \boldsymbol{H}, \boldsymbol{\tau}) = \frac{1}{M^K} \sum_{\boldsymbol{x}' \in \mathcal{M}^{2K}} p(\boldsymbol{r} \mid \boldsymbol{x} = \boldsymbol{x}', \boldsymbol{H}, \boldsymbol{\tau}).
\end{equation}

Now that the first term in (\ref{eqn:mi_def}) is found, we can move on to the second term. Due to the conditional independence of the quantized observations given the input vector, the channel matrix, and the quantization thresholds, we can write

\begin{equation}
\label{eqn:mi_second}
\mathcal{H}(\boldsymbol{r} \mid \boldsymbol{x}, \boldsymbol{H}, \boldsymbol{\tau}) =
\sum_{n=1}^{2N} \mathcal{H}(r_n \mid \boldsymbol{x}, \boldsymbol{H}, \tau_n) = \frac{1}{M^K} \sum_{n=1}^{2N} \sum_{\boldsymbol{x}' \in \mathcal{M}^{2K}} \mathcal{H}(r_n \mid \boldsymbol{x} = \boldsymbol{x}', \boldsymbol{H}, \tau_n).
\end{equation}

Again, by definition, and since each $r_n$ is a binary random variable,

\begin{equation}
\label{eqn:entropy_second}
\mathcal{H}(r_n \mid \boldsymbol{x}, \boldsymbol{H}, \tau_n) = 
\sum_{r_n \in \{\pm 1\}} p(\boldsymbol{r} \mid \boldsymbol{x}, \boldsymbol{H}, \boldsymbol{\tau}) \log_2 (p(\boldsymbol{r} \mid \boldsymbol{x}, \boldsymbol{H}, \boldsymbol{\tau})) = \mathcal{H}_b \left( \Phi \left( \frac{ \boldsymbol{h}_n^T \boldsymbol{x} - \tau_n}{\sqrt{N_0/2}} \right) \right).
\end{equation}

\bibliographystyle{IEEEtran}
% \balance
\bibliography{main}

\begin{thebibliography}{10}
\providecommand{\url}[1]{#1}
\csname url@samestyle\endcsname
\providecommand{\newblock}{\relax}
\providecommand{\bibinfo}[2]{#2}
\providecommand{\BIBentrySTDinterwordspacing}{\spaceskip=0pt\relax}
\providecommand{\BIBentryALTinterwordstretchfactor}{4}
\providecommand{\BIBentryALTinterwordspacing}{\spaceskip=\fontdimen2\font plus
\BIBentryALTinterwordstretchfactor\fontdimen3\font minus
  \fontdimen4\font\relax}
\providecommand{\BIBforeignlanguage}[2]{{%
\expandafter\ifx\csname l@#1\endcsname\relax
\typeout{** WARNING: IEEEtran.bst: No hyphenation pattern has been}%
\typeout{** loaded for the language `#1'. Using the pattern for}%
\typeout{** the default language instead.}%
\else
\language=\csname l@#1\endcsname
\fi
#2}}
\providecommand{\BIBdecl}{\relax}
\BIBdecl

\bibitem{Massive_MIMO_Survey}
F.~Jameel, Faisal, M.~A.~A. Haider, and A.~A. Butt, ``Massive {MIMO}: A survey
  of recent advances, research issues and future directions,'' in \emph{2017
  Int. Symp. Recent Advances in Elect. Eng. (RAEE)}, 2017, pp. 1--6.

\bibitem{Low_Res_ADC_Survey}
J.~Liu, Z.~Luo, and X.~Xiong, ``Low-resolution {ADC}s for wireless
  communication: A comprehensive survey,'' \emph{{IEEE} Access}, vol.~7, pp.
  91\,291--91\,324, 2019.

\bibitem{CE_and_Perf_Analysis}
Y.~Li, C.~Tao, G.~Seco-Granados, A.~Mezghani, A.~L. Swindlehurst, and L.~Liu,
  ``Channel estimation and performance analysis of one-bit massive {MIMO}
  systems,'' \emph{{IEEE} Trans. Signal Process.}, vol.~65, no.~15, pp.
  4075--4089, 2017.

\bibitem{Perf_Bound_Rayleigh_Channel}
T.-k. Kim, M.~Min, and Y.-S. Jeon, ``Performance bound for {MIMO} systems using
  one-bit {ADC}s over {Rayleigh} fading channels,'' \emph{{IEEE} Trans. Veh.
  Tech.}, pp. 1--1, 2022.

\bibitem{Throughput_Analysis}
S.~Jacobsson, G.~Durisi, M.~Coldrey, U.~Gustavsson, and C.~Studer, ``Throughput
  analysis of massive {MIMO} uplink with low-resolution {ADC}s,'' \emph{{IEEE}
  Trans. Wireless Commun.}, vol.~16, no.~6, pp. 4038--4051, 2017.

\bibitem{Wideband_1bit_Perf}
C.~Mollén, J.~Choi, E.~G. Larsson, and R.~W. Heath, ``Uplink performance of
  wideband massive {MIMO} with one-bit {ADC}s,'' \emph{{IEEE} Trans. Wireless
  Commun.}, vol.~16, no.~1, pp. 87--100, 2017.

\bibitem{Comm_Limits_Low_Res_ADC}
J.~Singh, O.~Dabeer, and U.~Madhow, ``On the limits of communication with
  low-precision analog-to-digital conversion at the receiver,'' \emph{{IEEE}
  Trans. Commun.}, vol.~57, no.~12, pp. 3629--3639, 2009.

\bibitem{CSIT_Capacity}
J.~Mo and R.~W. Heath, ``Capacity analysis of one-bit quantized {MIMO} systems
  with transmitter channel state information,'' \emph{{IEEE} Trans. Signal
  Process.}, vol.~63, no.~20, pp. 5498--5512, 2015.

\bibitem{Capacity_High_SNR_mmWave}
------, ``High {SNR} capacity of millimeter wave {MIMO} systems with one-bit
  quantization,'' in \emph{2014 Inf. Theory and Appl. Workshop (ITA)}, 2014,
  pp. 1--5.

\bibitem{sr_detectability}
M.~Chen, N.~Q. Hu, G.~J. Qin, and Y.~M. Yang, ``A study on
  additional-signal-enhanced stochastic resonance in detecting weak signals,''
  in \emph{2008 {IEEE} Int. Conf. Netw., Sens. and Control}, 2008, pp.
  1636--1640.

\bibitem{Stoch_Res_Review}
G.~Harmer, B.~Davis, and D.~Abbott, ``A review of stochastic resonance:
  circuits and measurement,'' \emph{{IEEE} Trans. Instrum. and Meas.}, vol.~51,
  no.~2, pp. 299--309, 2002.

\bibitem{sr_dynamic}
Q.~Ye, H.~Huang, X.~He, and C.~Zhang, ``A study on the parameters of bistable
  stochastic resonance systems and adaptive stochastic resonance,'' in
  \emph{{IEEE} Int. Conf. Robotics, Intell. Syst. and Signal Process.}, vol.~1,
  2003, pp. 484--488 vol.1.

\bibitem{Medical_Imaging_SR}
L.~Yue, P.~Ganesan, B.~S. Sathish, C.~Manikandan, A.~Niranjan, V.~Elamaran, and
  A.~F. Hussein, ``The importance of dithering technique revisited with
  biomedical images—a survey,'' \emph{{IEEE} Access}, vol.~7, pp. 3627--3634,
  2019.

\bibitem{Stoch_Res_as_Dither}
R.~Wannamaker, S.~Lipshitz, and J.~Vanderkooy, ``Stochastic resonance as
  dithering,'' \emph{Physical review. E, Statistical physics, plasmas, fluids,
  and related interdisciplinary topics}, vol.~61, pp. 233--6, Feb. 2000.

\bibitem{dafsp}
I.~Bilinskis, \emph{Digital Alias-Free Signal Process.}\hskip 1em plus 0.5em
  minus 0.4em\relax USA: John Wiley \& Sons, Inc., 2007.

\bibitem{Deep_Signal_Recov}
S.~Khobahi, N.~Naimipour, M.~Soltanalian, and Y.~C. Eldar, ``Deep signal
  recovery with one-bit quantization,'' in \emph{ICASSP 2019 {IEEE} Int. Conf.
  Acoust., Speech and Signal Process.}, 2019, pp. 2987--2991.

\bibitem{Generalized_Bussgang_CE}
Q.~Wan, J.~Fang, H.~Duan, Z.~Chen, and H.~Li, ``Generalized {Bussgang} {LMMSE}
  channel estimation for one-bit massive {MIMO} systems,'' \emph{{IEEE} Trans.
  Wireless Commun.}, vol.~19, no.~6, pp. 4234--4246, 2020.

\bibitem{Quant_Design_and_CE}
F.~Wang, J.~Fang, H.~Li, Z.~Chen, and S.~Li, ``One-bit quantization design and
  channel estimation for massive {MIMO} systems,'' \emph{{IEEE} Trans. Veh.
  Tech.}, vol.~67, no.~11, pp. 10\,921--10\,934, 2018.

\bibitem{Antithetic_Dither_CE}
D.~K.~W. Ho and B.~D. Rao, ``Antithetic dithered 1-bit massive {MIMO}
  architecture: Efficient channel estimation via parameter expansion and
  {PML},'' \emph{{IEEE} Trans. Signal Process.}, vol.~67, no.~9, pp.
  2291--2303, 2019.

\bibitem{Perf_of_FTSR_Sampling}
A.~B. Üçüncü and A.~{\"O}. Yılmaz, ``Performance analysis of faster than
  symbol rate sampling in 1-bit massive {MIMO} systems,'' in \emph{2017 {IEEE}
  Int. Conf. Commun. (ICC)}, 2017, pp. 1--6.

\bibitem{Oversampling_MIMO_OFDM}
A.~B. Üçüncü, E.~Björnson, H.~Johansson, A.~{\"O}. Yılmaz, and E.~G.
  Larsson, ``Performance analysis of quantized uplink massive {MIMO}-{OFDM}
  with oversampling under adjacent channel interference,'' \emph{{IEEE} Trans.
  Commun.}, vol.~68, no.~2, pp. 871--886, 2020.

\bibitem{ADMM}
{\"O}.~T. Demir and E.~Björnson, ``{ADMM}-based one-bit quantized signal
  detection for massive {MIMO} systems with hardware impairments,'' in
  \emph{ICASSP 2020 {IEEE} Int. Conf. Acoust., Speech and Signal Process.},
  2020, pp. 9120--9124.

\bibitem{OBMNet}
L.~V. Nguyen, A.~L. Swindlehurst, and D.~H.~N. Nguyen, ``Linear and deep neural
  network-based receivers for massive {MIMO} systems with one-bit {ADC}s,''
  \emph{{IEEE} Trans. Wireless Commun.}, vol.~20, no.~11, pp. 7333--7345, 2021.

\bibitem{Bayes_Optimal_CE_DD}
C.-K. Wen, C.-J. Wang, S.~Jin, K.-K. Wong, and P.~Ting, ``{Bayes}-optimal joint
  channel-and-data estimation for massive {MIMO} with low-precision {ADC}s,''
  \emph{{IEEE} Trans. Signal Process.}, vol.~64, no.~10, pp. 2541--2556, 2016.

\bibitem{Ungerboeck_Rec}
A.~B. Üçüncü, G.~M. Güvensen, and A.~{\"O}. Y{\i}lmaz, ``A reduced
  complexity {Ungerboeck} receiver for quantized wideband massive
  {SC}-{MIMO},'' \emph{{IEEE} Trans. Commun.}, vol.~69, no.~7, pp. 4921--4936,
  2021.

\bibitem{Threshold_Rec_Design_and_Strategies}
A.~Khalili, F.~Shirani, E.~Erkip, and Y.~C. Eldar, ``{MIMO} networks with
  one-bit {ADC}s: Receiver design and communication strategies,'' \emph{{IEEE}
  Trans. Commun.}, vol.~70, no.~3, pp. 1580--1594, 2022.

\bibitem{nML}
J.~Choi, J.~Mo, and R.~W. Heath, ``Near maximum-likelihood detector and channel
  estimator for uplink multiuser massive {MIMO} systems with one-bit {ADC}s,''
  \emph{{IEEE} Trans. Commun.}, vol.~64, no.~5, pp. 2005--2018, 2016.

\bibitem{1BOX}
S.~H. Mirfarshbafan, M.~Shabany, S.~A. Nezamalhosseini, and C.~Studer,
  ``Algorithm and {VLSI} design for 1-bit data detection in massive
  {MIMO}-{OFDM},'' \emph{{IEEE} Open J. of Circuits and Syst.}, vol.~1, pp.
  170--184, 2020.

\bibitem{LoRDNet}
S.~Khobahi, N.~Shlezinger, M.~Soltanalian, and Y.~C. Eldar, ``Lord-net:
  Unfolded deep detection network with low-resolution receivers,'' \emph{{IEEE}
  Trans. Signal Process.}, vol.~69, pp. 5651--5664, 2021.

\bibitem{SVM}
L.~V. Nguyen, A.~L. Swindlehurst, and D.~H.~N. Nguyen, ``{SVM}-based channel
  estimation and data detection for one-bit massive {MIMO} systems,''
  \emph{{IEEE} Trans. Signal Process.}, vol.~69, pp. 2086--2099, 2021.

\bibitem{Robust_DD_Reinf_Learning}
Y.-S. Jeon, N.~Lee, and H.~V. Poor, ``Robust data detection for {MIMO} systems
  with one-bit {ADC}s: A reinforcement learning approach,'' \emph{{IEEE} Trans.
  Wireless Commun.}, vol.~19, no.~3, pp. 1663--1676, 2020.

\bibitem{Supervised_Learning_Comm_Framework}
Y.-S. Jeon, S.-N. Hong, and N.~Lee, ``Supervised-learning-aided communication
  framework for {MIMO} systems with low-resolution {ADC}s,'' \emph{{IEEE}
  Trans. Veh. Tech.}, vol.~67, no.~8, pp. 7299--7313, 2018.

\bibitem{Hamm_Dist_Decod}
S.-N. Hong, S.~Kim, and N.~Lee, ``A weighted minimum distance decoding for
  uplink multiuser {MIMO} systems with low-resolution {ADC}s,'' \emph{{IEEE}
  Trans. Commun.}, vol.~66, no.~5, pp. 1912--1924, 2018.

\bibitem{sc_fde}
J.~Guerreiro, R.~Dinis, and P.~Montezuma, ``Low-complexity {SC}-{FDE}
  techniques for massive {{MIMO}} schemes with low-resolution {ADC}s,''
  \emph{{IEEE} Trans. Commun.}, vol.~67, no.~3, pp. 2368--2380, Mar. 2019.

\bibitem{Bussgang_Dec}
O.~T. Demir and E.~Bjornson, ``The {Bussgang} decomposition of nonlinear
  systems: Basic theory and {MIMO} extensions [lecture notes],'' \emph{{IEEE}
  Signal Process. Mag.}, vol.~38, no.~1, pp. 131--136, 2021.

\bibitem{Oversampling_MIMO}
A.~B. Üçüncü and A.~{\"O}. Yılmaz, ``Oversampling in one-bit quantized
  massive {MIMO} systems and performance analysis,'' \emph{{IEEE} Trans.
  Wireless Commun.}, vol.~17, no.~12, pp. 7952--7964, 2018.

\bibitem{ss_diversity}
Y.~Shang, D.~Wang, and X.-G. Xia, ``Signal space diversity techniques with fast
  decoding based on {MDS} codes,'' \emph{{IEEE} Trans. Commun.}, vol.~58,
  no.~9, pp. 2525--2536, 2010.

\bibitem{Newtons_Method}
\BIBentryALTinterwordspacing
A.~Galántai, ``The theory of {Newton's} method,'' \emph{J. of Comput. and
  Appl. Math.}, vol. 124, no.~1, pp. 25--44, 2000. [Online]. Available:
  \url{https://www.sciencedirect.com/science/article/pii/S0377042700004350}
\BIBentrySTDinterwordspacing

\bibitem{optimization_comparison}
J.~Liang, ``Gradient descent and {Newton's} method with backtracking line
  search in linear regression,'' in \emph{2021 2nd Int. Conf. Comput. and Data
  Sci. (CDS)}, 2021, pp. 394--397.

\bibitem{qnd}
G.~Yılmaz and A.~{\"O}. Yılmaz, ``Quasi-{Newton} detection in one-bit
  pseudo-randomly quantized wideband massive {MIMO} systems,'' Apr. 2023,
  submitted to {IEEE} Trans. Wireless Commun. and currently under review.

\bibitem{yilmaz_2023}
G.~Yılmaz, ``Pseudo-random quantization based detection in one-bit massive
  {{MIMO}} systems,'' M.S. Thesis, Middle East Technical University, Ankara,
  Turkey, Feb. 2023.

\end{thebibliography}

\end{document}